# UNIVERSITÄT LUZERN

KULTUR- UND SOZIALWISSENSCHAFTLICHE FAKULTÄT
DEKANAT

## Automatic generation of DRI Statements

Masterarbeit

zur Erlangung des

Mastergrades

der Kultur- und Sozialwissenschaftlichen Fakultät der

Universität Luzern

vorgelegt von

Kevin Maurice Flechtner

S23-453-475

Eingereicht am: 31.3.2025

Erstgutachter: Dr. Francesco Veri

Zweitgutachter: Prof. Dr. Alexander Trechsel

# I. Abstract

Assessing the quality of group deliberation is essential for improving our understanding of deliberative processes. The Deliberative Reason Index (DRI) offers a sophisticated metric for evaluating group reasoning, but its implementation has been constrained by the complex and time-consuming process of statement generation. This thesis introduces an innovative, automated approach to DRI statement generation that leverages advanced natural language processing (NLP) and large language models (LLMs) to substantially reduce the human effort involved in survey preparation. Key contributions are a systematic framework for automated DRI statement generation and a methodological innovation that significantly lowers the barrier to conducting comprehensive deliberative process assessments. In addition, the findings provide a replicable template for integrating generative artificial intelligence into social science research methodologies.

## II. Table of Contents

## III. List of Figures

# 1. Introduction

To deliberative theory, democracy, at its core, is not only a mechanism for aggregating preferences but a process for finding reasonable solutions to political problems through reasoned discourse among free and equal citizens (Benson, 2021; Cohen, 2002; Landemore, 2013). Measuring the quality of such discourses can help to assess their success and increase our understanding of deliberation (Niemeyer et al., 2024). While assessments of the process, e.g. Discourse Quality Index (Steenbergen et al., 2003), help measure the quality of a deliberation and impact measures assess the effects of deliberative processes on aspects such as political engagement (Fishkin et al., 2024) or group polarization (Caluwaerts et al., 2023; Grönlund et al., 2023), it's outcome measures such as the Deliberative Reason Index (DRI) (Niemeyer & Veri, 2022) that help to provide significant information on the effectiveness of deliberative processes. With DRI, we can evaluate deliberation by measuring the intersubjective consistency[1] between participants through Q-methodology, a research technique used in cognitive psychology that combines qualitative and quantitative methods to analyze subjective viewpoints, before and after a deliberation to get a sense of the level of meta-consensus the participants arrive at (ibid., Stephenson, 1953).

Despite its potential, the application of the DRI has been constrained by the requirement of labor-intensive discourse analysis. Statements for the DRI survey are traditionally extracted from lengthy qualitative interviews and large-scale analyses of the media landscape (Niemeyer & Veri, 2022). Therefore, collecting the necessary data requires extensive human labor. The statements are limited in their representativeness by the small number of qualitative interviews and the limited amount of media being conducted for the survey. Additionally, the DRI method lacks standardization, putting the quality of DRI results and their comparability at risk (ibid.). These limitations hinder the widespread adoption of the DRI, within and outside of the scientific context, and restrict its utility in systematic comparisons.

Advancements in artificial intelligence (AI) and machine learning (ML) present promising opportunities to address these challenges. The capabilities of AI and ML to process vast amounts of unstructured textual data can automate tasks that previously required significant human labor (Landemore, 2024; McKinney, 2024). In this paper we will outline how, by leveraging these technologies, it is possible to mitigate the limitations of the DRI in two fundamental ways. First, automation through AI and ML can significantly reduce the human labor required for data analysis since manual coding and analysis inherent in the traditional application of the DRI are very resource intensive. AI and ML algorithms can automate much of this work, such as identifying core statements and categorizing statements, thereby increasing efficiency and scalability. By leveraging generative AI to accelerate, simplify and standardize the process of DRI survey generation, our method also enables the application of the DRI in more diverse settings, facilitating comparative studies and longitudinal analyses. As a second improvement, AI and ML can expand access to a broader range of reasons and perspectives. The traditional reliance on human coders for qualitative interviews and large-scale media analyses limits the amount of data that can be processed and therefore diversity of perspectives taken into account, introducing the risk of human bias. By utilizing extensive corpora of news articles to approximate the public sphere it becomes feasible to access a more comprehensive and representative array of viewpoints. Our approach can help to achieve the deliberative democratic ideal of inclusivity and enhance the validity of the DRI by reflecting a much larger part of public discourse as it would be possible with a few human analysts.

---

[1] Intersubjective consistency is the proportional agreement between participants regarding both the considerations (reasons) and preferences (choices) relevant to a deliberative issue. See Chapter 2.3 for a more detailed explanation.

The aim of this thesis is to demonstrate how state-of-the-art AI and ML techniques can be applied to generate DRI surveys in order to streamline its application and enhance its utility in assessing deliberative processes. Specifically, this research presents a methodological framework that employs a multilingual embeddings model and vector similarity measures to categorize news articles which are then fed into an LLM to generate core statements, considerations and reasons from the most relevant articles. This approach not only addresses the limitations of traditional methods but also improves the representational power of DRI. Methodologically, it advances the integration of AI and ML into political science research, demonstrating the practical application of advanced computational tools to complex social phenomena. By bridging qualitative and quantitative research methods, it offers an innovative approach to analyzing deliberative processes. The use of extensive text corpora and automated analysis enhances the depth and breadth of data captured, leading to more accurate and comprehensive assessments of deliberation. The massive increase of dataset size allows us to capture aspects of the public debate that might be missed in traditional qualitative interviews, thus improving the validity and reliability of findings. Furthermore, the methodologies developed in this thesis can be adapted and applied to the analysis of large-scale text corpora in other contexts, providing a foundation for future research in computational social sciences.

To illustrate the practical application of this methodology, the thesis focuses on the topic of health care costs in Switzerland, a subject of significant public interest and policy relevance and the most significant worry of the Swiss population according to the 'UBS Sorgenbarometer 2024' (Golder et al., 2024). The research supports the 'Bevölkerungsrat 2025', a citizens' assembly dedicated to addressing health care costs in Switzerland (Kübler & Stojanovic, 2022 - 2026). By applying our pipeline to the generation of a DRI survey for this case study, we aim to demonstrate the feasibility and practicability of our method. Furthermore, the method has been validated by application to the topic of food policy where we compared the results to an existing, human-made DRI survey produced for the Swiss Citizens' Assembly for Food Policy in 2022 (Mathys, 2024).

The structure of the thesis is organized as follows. Chapter 2 provides necessary context through an introduction into the relevant literature, exploring the theoretical foundations of deliberative democracy. Here we will introduce the most common existing measures of deliberative processes and review current applications of AI and ML in similar contexts, thereby identifying the research gaps that this thesis aims to address. At the end of the chapter, the 'Bevölkerungsrat 2025', the testbed for our method, will be presented. Chapter 3 explains our method in detail, including data collection, preprocessing, dimensionality reduction, categorization techniques and human assessment of the results. It outlines the use of AI and ML algorithms for generating DRI statements and policy options and discusses challenges encountered during the process. Chapter 4 presents a discussion of our findings, analyzing and evaluating the results, addressing the limitations of the method, and exploring its implications for digital democracy. It then offers recommendations for future research based on the findings. Finally, Chapter 5 sums up the key findings, highlighting the contributions to political science and digital democracy, and providing concluding thoughts on the broader relevance of the research.

## 2. Literature Review & Context

### 2.1 Deliberative Democracy

The field of Deliberative democracy emerged in the 20th century in response to the perceived limitations of aggregative and representative theories of democracy (Chambers, 2018; Mackie, 2018). Unlike these models, which emphasize voting and preference aggregation, deliberative democracy prioritizes reasoned dialogue, freedom and equality of all participants and collective decision-making as the foundational aspects of democracy (Cohen, 2002). In this chapter, we will first introduce the theoretical cornerstones of deliberative democracy, laid by Habermas in the continental tradition and Rawls & Cohen in the Anglo-American tradition. Then we will focus on questions of legitimacy, the ideal outcome of deliberative processes and the practical offsprings of deliberative theory in distinct processes and deliberative systems.

One of the first and most prominent contributors to deliberative theory is Habermas. His theory of communicative action emphasized the role of rational-critical debate in democratic decision-making. Habermas first laid the groundwork for his deliberative theory with his habilitation thesis *The Structural Transformation of the Public Sphere* (1991) in 1962. Here he introduced the concept of the public sphere as an inclusive space for reasoned debate, free from coercion. Later, he proposed *The Theory of Communicative Action* (1984), which focused on mutual understanding rather than instrumental or opportunistic use of rationality. In this model, individuals engage in argumentation with the aim of reaching consensus based on rational debate rather than coercion or manipulation.

In the Anglo-American tradition, Rawls played an important role in shaping deliberative democratic theory. His concept of public reason, introduced in *A Theory of Justice* (Rawls, 1971) and later developed in *Political Liberalism* (Rawls, 1993), emphasized the need for justifications in political discourse that all reasonable citizens can accept. Rawlsian deliberation focuses on fairness and the construction of overlapping consensus, ensuring that political principles are justified across diverse moral and philosophical perspectives rather than requiring absolute rational consensus. Expanding on Rawls, Cohen introduced an ideal deliberative procedure, in which free and equal citizens engage in collective reasoning guided by the common good (Cohen, 2002). Cohen also introduced an epistemic dimension, arguing that deliberation is not only about fairness but also about producing better, truth-tracking decisions (Cohen, 1986).

Questions on legitimacy have played a significant role in the theory of deliberative democracy since its early days. Manin argued that "[...] the source of legitimacy is not the predetermined will of individuals, but rather the process of its formation, that is, deliberation itself." (Manin, 1987, 352). Similar arguments were later put forward by Habermas and Cohen. In *Between Facts and Norms* (1996), Habermas introduces his discourse theory of democracy. Here he proposes the idea that legitimacy in democracy arises from discursive processes where citizens engage in rational argumentation under ideal speech conditions. In the same year, Cohen argued that the exercise of collective political power should be justified on the basis "[...] of a free public reasoning among equals." (1996, 99).

Moving beyond the rationalistic beginnings of deliberative theory, Pluralists such as Young (2000) and Sanders (1997) have pivoted away from the ideal of solely rational debate emphasizing that deliberation should include diverse forms of communication, including storytelling and rhetoric, to ensure inclusivity and to recognize the lived experiences of diverse groups with varying backgrounds. Agonists like Mouffe (1999) even argue that the inherent conflicts between interests, values and ways of living lie at the core of democracy and should be embraced in a respectful manner instead of mitigated. As a response to a widening criticism of consensus as a deliberative ideal, Niemeyer and Dryzek introduced the concepts of meta-consensus and intersubjective rationality (Bächtiger et al., 2007; Dryzek & Niemeyer, 2006; Niemeyer & Dryzek, 2007). Meta-consensus shifts away from the assumption that the goal of deliberation is to produce complete agreement on a given topic. Instead, a good deliberation should foster a shared understanding of the debate and its relevant considerations. According to this concept, good deliberation helps with identifying a set of relevant and legitimate reasons and values that can be used for decision-making, allowing participants to disagree on specific aspects while recognizing their collective importance. Niemeyer and Dryzek (2007) distinguish three types of meta-consensus: the identification of a common set of legitimate values (normative meta-consensus), credible reasons or beliefs relevant to the decision (epistemic meta-consensus), and an understanding of the acceptable options or alternatives that are open to consideration (preference meta-consensus). This approach moves beyond the Habermasian ideal of rational consensus by accommodating diverse perspectives and opinions while still maintaining normative aspirations for democratic legitimacy. The concept of meta-consensus will also build the basis of the DRI which will play a key role in our research paper.

With the concept of deliberative polling, Fishkin (1991) was the first to bring the theoretical ideas of deliberative democracy into practice. Starting from there, the development of practical applications of deliberative theory has played an increasingly important role. Scholars and policymakers have experimented with diverse deliberative formats such as deliberative mini-publics, citizens' assemblies, citizens' juries and participatory budgeting (Veri, 2023). These initiatives have been implemented in various democratic contexts, including local governance, national policymaking, and even constitutional reforms. Following the ideals of deliberative theory, these experiments aim to create spaces where citizens can engage in informed, reflective, and reasoned discussion to provide recommendations on policy matters.

Scholars like Mansbridge (1999) and Dryzek (2002) further expanded deliberative theory by highlighting the importance of deliberation in non-traditional spaces, such as transnational governance and informal public spheres, advocating for a more flexible and pluralistic approach. With the release of *Deliberative Systems* in 2012, this development became known as the deliberative systems approach (Parkinson & Mansbridge, 2012), which moves beyond single-site deliberation to recognize that democratic deliberation occurs across multiple, interconnected spaces. These include formal institutions such as legislatures, courts, and administrative bodies, as well as informal arenas like civil society organizations, public media, and everyday discussions. The systemic perspective acknowledges that no single site can fully embody the ideal conditions of deliberation, but that different components such as reason-giving in legislatures, accountability in the judiciary, and agenda-setting in the public sphere contribute in complementary ways to democratic legitimacy (Mansbridge et al., 2012; Owen & Smith, 2015). By distributing deliberation across a broader landscape, the deliberative systems approach enhances inclusivity, responsiveness, and real-world applicability, addressing concerns about power imbalances and the limitations of ideal speech conditions. It also mitigates the contrast between aggregative and deliberative theories of democracy. While early models of deliberative democracy positioned deliberation and aggregation as competing paradigms (Elster, 2002), more recent research suggests they are complementary, recognizing that deliberation can inform and improve decision-making in electoral and representative processes (Estlund & Landemore, 2018).

Deliberative democracy has undergone significant theoretical and empirical developments since its emergence. While early models emphasized rational debate and consensus, contemporary approaches recognize the importance of pluralism, inclusion, and systemic deliberation. The field continues to evolve, integrating insights from empirical research and adapting to the challenges posed by political polarization and digitalization.

## 2.2 Assessing Deliberation

As the theory of deliberative democracy continues to evolve and more and more practical applications of the theory are being developed, attention has turned to how the quality and impact of deliberative processes could possibly be measured. A key distinction in this regard is the distinction between measuring the conditions that should theoretically enable deliberation, such as equality, inclusivity, and access to information, and measuring the outcomes of deliberation, such as shifts in opinion, enhanced mutual understanding, or the quality of policy decisions. Most available measurements focus on the procedural aspect of deliberation, trying to operationalize and evaluate conditions that foster meaningful exchange. Those attempts at systematic evaluation typically begin by defining what constitutes a 'good' deliberative process. This can be achieved by drawing upon normative theories of deliberative democracy and translating them into empirically observable indicators.

One of the most prominent tools in this domain is the Discourse Quality Index (DQI) (Steenbergen et al., 2003), which is grounded in Habermasian discourse ethics (Habermas, 1990) and has been widely applied across diverse empirical contexts (Elstub et al., 2022; Lord & Tamvaki, 2013). DQI measures the quality of deliberation on five dimensions: participation, level of justification, content of justifications, respect and constructive politics (Steenbergen et al., 2003). Over time, the methodological literature on deliberation has expanded upon the DQI's initial framework, improving upon the measurement of equality, the degree of interactivity and reflection within the discussion, and the substantive argumentative quality and sincerity of contributions (Bächtiger et al., 2022).

Additional measurement tools have emerged that target more specific aspects of deliberative processes. The Listening Quality Index (Scudder, 2022), for instance, focuses on the quality of listening behaviors in deliberative settings. The Online Deliberative Matrix (Kies, 2022) is explicitly designed to evaluate deliberation in online settings. Drawing on Habermasian discourse theory, it examines inclusion, reciprocity, empathy, sincerity, and the external impact of online discussions. Wyss et al. (2015) developed an approach to measure cognitive complexity as a proxy for deliberative quality. Cognitive complexity reflects individuals' abilities to perceive and integrate multiple dimensions of a topic, tolerate uncertainty, and work toward consensus.

Scholars have also developed methods to evaluate the impact of deliberative processes on participants, examining how participation affects factors such as political engagement, group polarization, trust, knowledge acquisition, civic empowerment, and attitude change. One common approach is to administer pre- and post-deliberation surveys to detect shifts in participants' attitudes such as political and social trust or political efficacy (Strandberg et al., 2021). By comparing opinions before and after deliberation, researchers can assess whether discussion moderates or reinforces group polarization. For instance, a review focussing on polarization and deliberation found deliberation studies report positive effects on participants' polarization levels, when deliberative ideals are adhered (Caluwaerts et al., 2023). Additionally, mini-publics often impact participants' political attitudes, knowledge, internal efficacy and reasoning skills positively, as shown in a recent meta-analysis (Theuwis et al., 2024). Beyond immediate post-test measures, longitudinal panel studies follow deliberation participants over time, often including a non-deliberating control group, to see if initial changes persist and translate into real-world behavior. For example, Fishkin et al. (2024) tracked citizens who took part in a national deliberative forum and found sustained increases in political engagement, including higher voter turnout one year later among deliberators compared to controls. Researchers have also used indicators such as observing subsequent voting records, civic participation, or community involvement to quantify deliberation's impact on engagement in practice (ibid.).

Apart from measuring the impacts of deliberation on participants, researchers also try to measure the quality of the outcome of such a process. This promptly poses the question of what a good outcome of a deliberation is. Early theorists grounded in the Habermasian theory of communicative action were particularly drawn to the notion that consensus should be the ultimate goal of deliberation. Believing in the "unforced force of the better argument" (Habermas, 1996, 306), they would argue that, under ideal conditions, participants of a deliberation should naturally converge to consensus during the process. In other words, consensus, which could also be understood as an alignment of rational wills, becomes an indicator that deliberation has successfully revealed and elevated the better argument.

However, the initial consensus-based approach to measuring outcomes has been subject to extensive critique and, in many respects, widely discredited. It has been argued that the goal of consensus could be both undesirable (Sanders, 1997; Young, 2002) or even unachievable (Femia, 1996; van Mill, 1996) in the first place. Aiming for complete consensus is unrealistic in political settings, where participants' values, interests, and interpretations of facts vary considerably. Not only is absolute unanimity rare, but its pursuit may also be questionable from a normative point of view, as it risks marginalizing minority voices or pressuring participants toward conformity at the expense of genuine engagement. Furthermore, the empirical challenge of determining whether consensus reflects genuinely reasoned agreement or merely a submission to the majority's view, renders consensus a problematic metric (Dryzek & Niemeyer, 2006).

Critiques of the goal of consensus as an outcome of deliberation have led to considerations of alternative approaches to capturing deliberation quality through outputs. On the theoretical side, Estlund and Landemore (2018) argued that deliberation has a unique epistemic quality which they summarize in the statement "diversity trumps ability" (ibid.), meaning that truth will be distilled through the process of a large group of people actively discussing the issue at hand. One could conclude that deliberation should yield substantially better, more 'correct' decisions, than e.g. the aggregated sum of individual decisions. Deliberation would thereby serve as a mechanism for producing outcomes closer to an independent standard of truth or goodness. However, this notion faces its own methodological and conceptual problems. By placing the benchmark for quality outside the deliberative process and the stakeholder's judgement, epistemic quality metrics risk becoming paternalistically detached from the deliberative procedures themselves. Moreover, they may undermine procedural legitimacy and the core value of inclusivity by referring to expert judgments on, or predefined definitions of a 'good' outcome (Chambers, 2017).

The critiques of outcome-based measures have motivated the search for more nuanced approaches that could account for both the procedural integrity and the unpredictable character of the outputs of deliberative processes (Dryzek & Niemeyer, 2006; Niemeyer & Dryzek, 2007). As a response, scholars have developed a measure known as the DRI, which attempts to assess the outcomes of deliberative processes on their own terms.

## 2.3 The Deliberative Reason Index

The DRI is a versatile outcome measure for deliberative processes that does not impose any strong external expectations on the process (Niemeyer & Veri, 2022). It builds on a rich theoretical foundation that integrates concepts from deliberative democratic theory, philosophy and cognitive sciences to assess the coherence and rationality of collective decision-making processes (ibid.). Central to DRI is a concept of 'reason' developed by Mercier and Sperber (2011). Here, arguments are considered to have the function of linking premises and conclusions. Reasoning, in this conceptualization, serves as a biological and social function to build trust and coherence among agents by enabling them both to produce arguments and to evaluate those offered by others. This inherently dialogical view of reasoning lends itself to the measurement of deliberation's quality, since the strength of a deliberative process is traditionally expected to lie in the exchange of reasons.

With reasoning being the main act of deliberation, DRI defines two outcomes of an ideal deliberative process. Firstly, Meta-consensus, which is an "[…] agreement about the nature of the issue at hand, not necessarily on the actual outcome." (Niemeyer & Dryzek, 2007). Meta-consensus arises when participants of a deliberation transcend their private interests and engage in a debate that incorporates all relevant arguments for the issue at hand. This in turn creates an agreement on the set of relevant values, preferences and possible implications the preferences might have which is meta-consensus (Dryzek & Niemeyer, 2006). Meta-consensus does not require individuals to agree to the same ideas, they only need to accept that all ideas within the set are relevant to the debate (Niemeyer & Dryzek, 2007). Secondly, intersubjective rationality conceptualizes how consistent individuals are in their reasoning (ibid.). An intersubjectively rational group of people is one where members of the group that agree on preferences also agree on relevant reasons and values for these preferences. Conversely, members of the group that disagree on preferences should also disagree on the relevant values and reasons for these preferences (ibid.). Therefore, the group is not required to come to a consensus to be intersubjectively consistent, it is merely required to agree that certain opinions and values lead to specific preferences while others lead to other preferences. Taken together, meta-consensus and intersubjective rationality move beyond simplistic measures of consensus for deliberation outcomes. The conceptual structure of meta-consensus and intersubjective consistency allows the DRI to overcome previous critiques by focusing on the underlying rational structure of deliberative reasoning and the coherence of beliefs, values and preferences (Dryzek & Niemeyer, 2006; Niemeyer & Dryzek, 2007).

The empirical application of the DRI relies on Q-methodology (Stephenson, 1953). Q-methodology is being used within the DRI survey to identify and categorize participants' belief systems and value orientations, based on how they rate various statements related to the deliberated issue. In the second part of the DRI survey, participants also express their preferences over concrete policy options, typically through a ranking task. Gathering this data at multiple time points, at least before and after a deliberative process, allows researchers to infer on the reasoning quality of deliberation. Reasoning quality is measured by correlating each participant's responses with those of other participants, separately evaluating agreement on considerations (reasons) and on preferences (action positions or policy choices) (Niemeyer & Veri, 2022). This approach captures whether shared considerations among participants translate coherently into collective preferences. Intersubjective consistency, which is central to the concept of DRI, refers to the degree of proportional agreement on both considerations and preferences among pairs of participants. By separately assessing agreement on considerations and preferences, the DRI reveals whether a group's collective reasoning aligns with their stated preferences, effectively capturing both the quality and consistency of deliberative reasoning (ibid.). DRI survey results are usually visualized as a scatter plot in a cartesian coordinate system spanning the two dimensions of intersubjective agreement on considerations on the x-axis and intersubjective agreement on preferences on the y-axis[2]. In such a visualization, each point corresponds to one pair of participants and complete intersubjective consistency corresponds to all points on the line $x = y$. The closer the points are to the $x = y$ axis, the higher the DRI score, meaning higher intersubjective consistency.

A typical DRI study begins by designing a survey instrument informed by preliminary qualitative interviews, media analyses and stakeholder consultations to identify the most salient beliefs, values, considerations and policy options in the public debate on the issue at hand. Such a process can be extremely time- and labor-intensive. Capturing the public debate on the issue at hand is crucial to the whole process, since the final DRI score will only include aspects that appear in the DRI survey. With DRI measuring the correlation between considerations and preferences, it is of particular importance that both sets are complete in order to prevent any case where the DRI survey lacks considerations or preferences from certain participants. Such a lack could lead to unexplainable diversions in agreement on considerations or preferences without any change in the other category.

Beliefs, values and considerations are usually summarized as consideration statements where agreement can be rated on a Likert-scale (Niemeyer & Veri, 2022). Policy options are listed separately as preferences and can usually be ranked from most to least preferred. Participants are encouraged to reflect deeply before answering, ensuring that their responses are grounded in genuine engagement with the material and conversations they have encountered during the deliberation. The survey is administered before and after the deliberative process to be able to assess differences in the answer patterns and infer influences from the process.

---

[2] See Figure 1 for an example DRI plot.

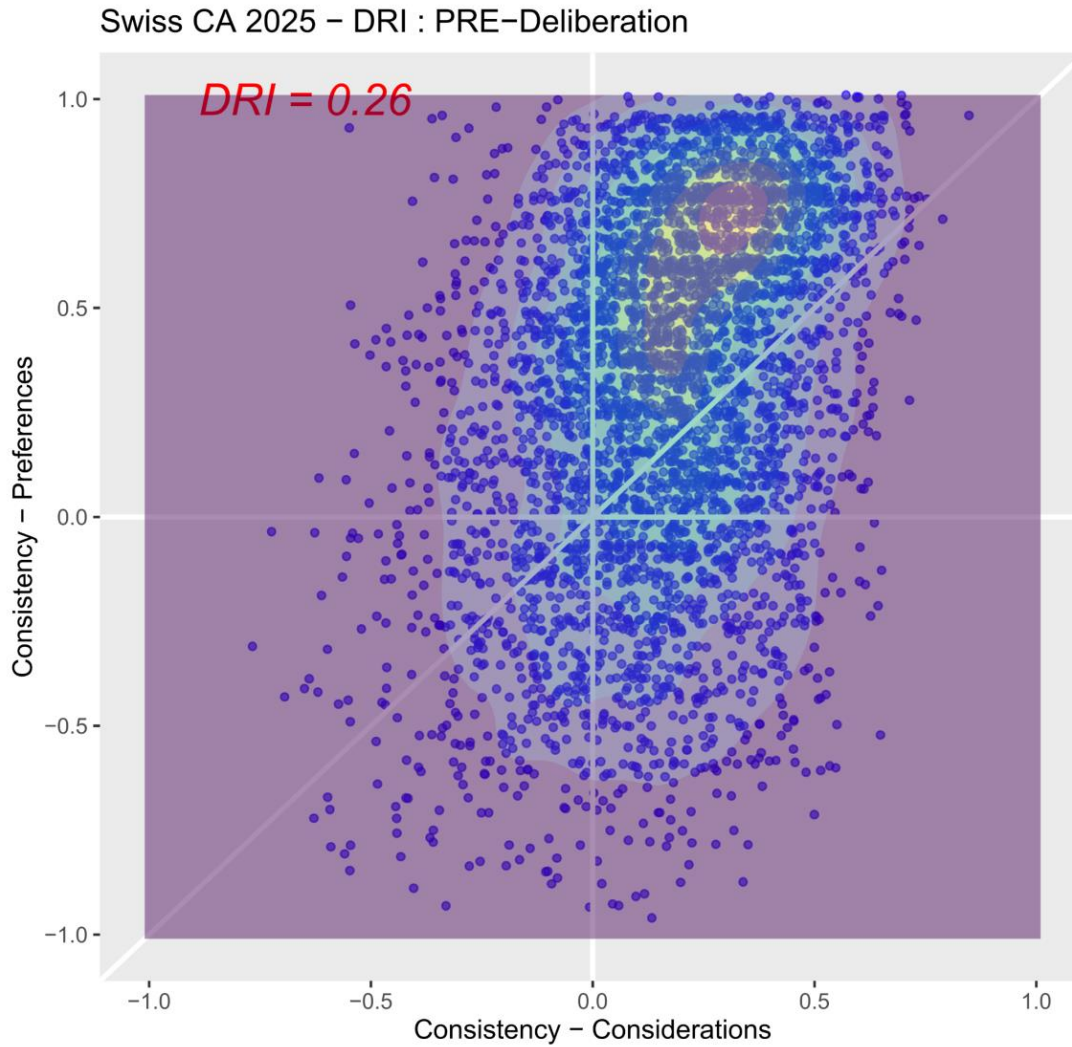

*Figure 1: The Pre-Deliberation DRI Plot of the 'Bevölkerungsrat 2025'*

Once the data collection is complete, four steps are applied to measure the final DRI score (ibid.). First, all responses on considerations and preferences are correlated pairwise using Spearman correlation. Then the pairwise level of consistency is calculated by calculating the distance to the intersubjective consistency line. Afterwards, individual DRI is calculated for each participant by averaging the distance to the x = y line for all pairs including the given individual. Lastly, the individual DRI scores are aggregated to group level by averaging over all individual DRI scores (ibid.). DRI is directly negatively correlated to the average distance from the x = y line for all pairwise combinations of participants, meaning DRI goes up when the average distance goes down. Having calculated DRI scores from pre- and post-deliberation surveys, one can then infer on the impact of the deliberative process on intersubjective consistency by comparing both DRI scores.

The DRI offers both a theoretically sound and methodologically stable tool to examine the impacts of deliberation. It builds on a strong philosophical foundation about the nature of reasoning and the impact of deliberation to create an inherently stable measure for concrete, empirical analysis of real-world deliberative processes. The DRI tries to refrain from imposing any outside value judgements on the outcomes of deliberation and instead provides a tool that, if applied correctly, allows to analyze the quality of deliberative reason while staying indifferent to the content of the deliberation (Niemeyer et al., 2024).

## 2.4 Applications of ML & AI Methods in Deliberation Assessments

In recent years, research has begun to apply ML and AI to improve and automate the assessment of deliberative processes. While no attempts have yet been made to automate the generation of DRI statements or the measurement of DRI in general, several initiatives have focused on automating the assessment of DQI. In this subchapter we will look at some of the latest and most prominent examples of automated DQI adaptations like DelibAnalysis (Fournier-Tombs & Di Marzo Serugendo, 2019) and AQuA (Additive deliberative Quality score with Adapters) (Behrendt et al., 2024) to outline the state of research in the field of automated deliberation assessment.

DelibAnalysis (Fournier-Tombs & Di Marzo Serugendo, 2019) employs ML to evaluate online political discourse. It adapts the general idea of DQI, using large, human-annotated datasets to train random forest classifiers to predict the deliberative quality of discussions at scale (ibid.). This method identifies features such as justification, respect, and the presence of counterarguments. It is employed to analyze large textual corpora, possibly generating insights into how context, moderation, or specific communication channels affect deliberative quality. Although promising, DelibAnalysis still relies heavily on manually labeled training data and uses simplified versions of the DQI criteria.

The AQuA framework (Behrendt et al., 2024) integrates pre-trained transformer language models alongside adapter models fine-tuned for specific deliberative aspects. Similarly to DQI, the AQuA approach disaggregates deliberative quality into multiple criteria, such as rationality, respect, and storytelling, and trains specialized adapters to predict scores for each. These scores are then weighted according to their correlation with expert assessments, producing a single integrated deliberative quality measure. The method begins to bridge the gap between expert-coded criteria and automated scoring, demonstrating how AI can capture complex, multi-dimensional constructs of discourse quality.

Other research focuses on narrower sets of deliberative dimensions. For instance, the work by Falk and Lapesa (2022) explores four key aspects of deliberative quality justification, common good orientation, interactivity, and respect. It uses both feature-based classifiers such as gradient-boosted trees and transformer models like RoBERTa (A Robustly Optimized BERT Pretraining Approach) (Liu et al., 2019) which is based on BERT (Bidirectional Encoder Representations from Transformers) (Devlin et al., 2019). While simpler feature-based models use handcrafted linguistic and sentiment features to achieve moderate performance, the transformer-based models generally surpass them, performing reliably even in low-resource scenarios. To mitigate issues like class imbalance, data augmentation techniques are employed, improving the classification of rarer deliberative qualities. Although integrating measures of argument quality, like cogency or reasonableness, yields only modest gains, this research underscores the potential of advanced natural language processing (NLP) techniques to identify subtle, deliberation-relevant signals in textual data.

Beyond text classification methods, some scholars have also harnessed network analysis and time-series modeling to capture the dynamics of online deliberation. The work by Shin and Rask (2021) exemplifies this shift, introducing 'throughput' indicators such as participation rate, activeness, continuity, responsiveness, inter-linkedness, and commitment. It applies them to Helsinki's OmaStadi budgeting process, a large-scale participatory initiative. By examining two-node-set-networks linking proposals and actors and employing time-series analysis to track fluctuations in engagement, their approach reveals how deliberation unfolds over time and how participants' interactions shape the deliberative space.

As an honorable mention, Gold et al. (2017) combine automated linguistic annotation with visual analytics to evaluate deliberative quality. By examining participation, respect, argumentation and justification, and persuasiveness, their methodology identifies linguistic features, such as causal connectors, discourse markers, and rhetorical devices. It then traces these patterns of deliberation over the course of real-world debates. Combining these automated annotations with interactive visual tools allows researchers and practitioners to explore how topical structure, speaker dominance, and interruptions evolve, offering nuanced insights into complex discourse processes. Even though it was an impressive leap forward in the automated analysis of deliberation at its time, the lexical approach quickly arrives at limitations of context compared to more recent applications of transformer-based models.

Collectively, these projects illustrate the expanding landscape of AI and ML applications in deliberation research. While existing methods have only focused on classification tasks relevant to the adaptation of the DQI to digital environments, one can still make out a general trend towards more automated forms of assessment. The refinement of computational approaches, the development of more sophisticated models, and the aggregation of more and more data suggest that automated deliberation assessment is looking at significant advancements in the upcoming years. As research discovers more complex applications of AI and ML for deliberative practice (Tessler et al., 2024), these computational techniques can also be applied to more complex measures like DRI, enabling scalable, data-driven evaluations of deliberation that maintain theoretical richness and nuanced interpretation.

## 2.5 Synthesis & Research Gaps

The existing literature on measuring deliberative quality reveals a diverse array of approaches, ranging from normative frameworks over procedural indicators to outcome-based measures that aim to capture shifts in beliefs and reasoning. Methodological advances, including the DRI, have addressed some of the theoretical and empirical shortcomings identified in earlier consensus-based metrics. Yet, significant gaps remain, particularly at the intersection of human-coded deliberative measures and the emerging capabilities offered by ML and AI. While applications of ML and AI to deliberation have begun to appear, these efforts mainly focus on classification tasks. Recent studies that attempt to automate the measurement of DQI or to scale up the assessment of deliberative content generally rely on rudimentary ML methods. Although these initiatives made it possible to handle larger datasets and to examine discourse more dynamically, their reliance on training datasets and the algorithms' limited understanding of context constrains their applicability and reduces their granularity.

One of the reasons why there is much more research on automated DQI measures might also be the complex nature of the DRI measure. Trying to measure intersubjectivity, DRI requires participants to answer a DRI survey tailor made for each use case. ML models designed for classification are not capable of solving such generative tasks. Only with the recent improvements in generative AI methods, especially LLMs, is it possible to automate such tasks. Generative AI models are widely expected to bring significant changes to deliberative practice (Landemore, 2024; Linegar et al., 2023; Mikhaylovskaya, 2024; Sleigh et al., 2024) and have already led to some promising innovation in the field (Tessler et al., 2024). Here, Tessler et al. present a method to iteratively summarize participant statements with an LLM to produce a statement that all participants agree on. However, the integration of generative models into deliberation measurement remains at an early stage, focusing only on the application of small transformer based models from the BERT family for classification tasks (Behrendt et al., 2024; Falk & Lapesa, 2022). At present, no automated approaches exist for generating the statements essential to DRI measurement or for extracting key statements representational for public discourse from large datasets in general. In particular, there has been no sustained effort to leverage generative AI and advanced NLP for generating key political statements directly from text, leaving the research gap that this paper tries to fill.

With the process of capturing the public debate on a given topic being extremely labor intensive and mistakes potentially leading to large gaps in representativeness of the survey, an automated approach to DRI survey generation would be particularly useful and essential for a more automated and standardized assessment of DRI. By outlining how to apply the generative capabilities of LLMs while maintaining strict control over the input data to control the outputs, we outline a useful application of the new capabilities of LLMs that could also be widely applied in computational social sciences for the extraction of key statements representational of a large discourse.

## 2.6. Case Study: The 'Bevölkerungsrat 2025' Citizens' Assembly

The empirical testbed for this automated approach to generating DRI statements introduced in this thesis is the 'Bevölkerungsrat 2025', a citizens' assembly on the topic of health costs in Switzerland. The initiative draws together the University of Zurich and the University of Geneva under the coordination of the Zentrum für Demokratie Aarau. Funding is primarily provided by the BRIDGE Discovery Project, supplemented by contributions from the Stiftung Mercator Schweiz.

The central aim of the citizens' assembly is to explore complementary forms of deliberative democracy and their potential to foster more meaningful public debate, mitigate political polarization, and ultimately influence policy-making and broader societal discourse. To ensure diversity and broad representation, a two-stage random selection process was put into place, initially contacting 22,000 individuals drawn from the Swiss population. From this initial pool, a stratified final sample of 100 participants was selected at random, reflecting a cross-section of the citizenry with respect to age, gender, education, political orientation, and regional backgrounds.

Running between November 2024 and March 2025, the deliberative format of the 'Bevölkerungsrat 2025' involved a series of encounters blending in-person and online sessions. Three weekend meetings are scheduled to take place in Zurich, Neuchâtel, and Bern, complemented by four online gatherings on Tuesday evenings. The deliberative process is professionally facilitated to encourage balanced, respectful, and constructive dialogue. Participants are not required to prepare in advance. Instead, they receive all necessary information and materials on-site. This ensures that the deliberation is accessible to all, regardless of prior knowledge or political experience.

In the selection of topics, the assembly places particular emphasis on issues that resonate with the broader public. The most pressing issue at the time, chosen by over 40% of respondents in preliminary surveys, centers on the challenge of rising healthcare costs. The final report generated by the assembly will provide policymakers, stakeholders, and the general public with informed, reasoned input on healthcare cost management.

Beyond the immediate policy suggestions, the 'Bevölkerungsrat 2025' has broader research and societal objectives. Academically, it offers a rare opportunity to study the effects of a large-scale, randomly selected deliberative body on public opinion, social cohesion, and the quality of democratic processes. The initiative thus tests the viability and legitimacy of citizens' assemblies in a complex policy environment, offering lessons that may inform the design of future deliberative forums and participatory structures. Additionally, public engagement is encouraged beyond the assembly's 100 members, with open invitations for input through online platforms.

## 3. Methodology

### 3.1 Method Overview

Traditionally, the generation of DRI statements has been a labor- and time-intensive task, often reliant on qualitative interviews and extensive manual analysis (Niemeyer & Veri, 2022). This chapter introduces a method to automate the process of DRI statement generation and thereby enhancing representativeness, efficiency and scalability. The proposed method is loosely related to Retrieval-Augmented Generation (RAG). Introduced by Lewis et al. (2021), RAG is a technique widely used to enhance the accuracy of generative models (Gao et al., 2024; Gupta et al., 2024). It does so by providing the generation model with relevant background information retrieved from a vector database. In our case, the model's outputs consist of DRI statements and policy proposals, informed by the most relevant paragraphs from news articles on the given topic. Our method encompasses six steps to achieve its goal[3]. First, a relevant dataset of text data related to the topic at hand needs to be assembled. In our case, we collected Swiss news articles using the Swissdox API (Application Programming Interface). This extensive corpus then serves as a proxy for public discourse, capturing a wide range of viewpoints. Second, anchor definitions are created for the predefined subcategories relevant to the debate with the help of an LLM. These categories provide a structured framework for the analysis of the corpus and can include both general themes and specific issues within the broader topic. The anchor definitions will serve as reference points for categorizing and interpreting the textual data later. Third, an encoder-only language model is applied to transform both the paragraphs and the anchor definitions into high-dimensional vector representations, called embeddings. This embedding process captures the semantic essence of the text, allowing for computational manipulation and analysis of linguistic content across the dataset. Fourth, the semantic similarity between each paragraph and the anchor definitions is calculated using cosine similarity measures. This step quantitatively assesses the relevance of each paragraph to the predefined subcategories, facilitating the identification of the most relevant pieces of text within the vast corpus. Fifth, for each grouping, the top five hundred paragraphs with the highest similarity scores are selected. These paragraphs are then attached to the input to a Large Language Model which is prompted to generate DRI statements and policy proposals from the data. In our case, this automated generation process yielded a total of 200 statements and 25 policy proposals, encompassing general themes, specific subcategories, and policy options across different political perspectives. Finally, the generated statements undergo human evaluation. The evaluation team reviews each statement for relevance, clarity, and lack of redundancy. This human oversight ensures the quality and validity of the outputs, combining the efficiency of automation with the discernment of expert judgment.

---

[3] See Figure 2 for an illustration of the steps.

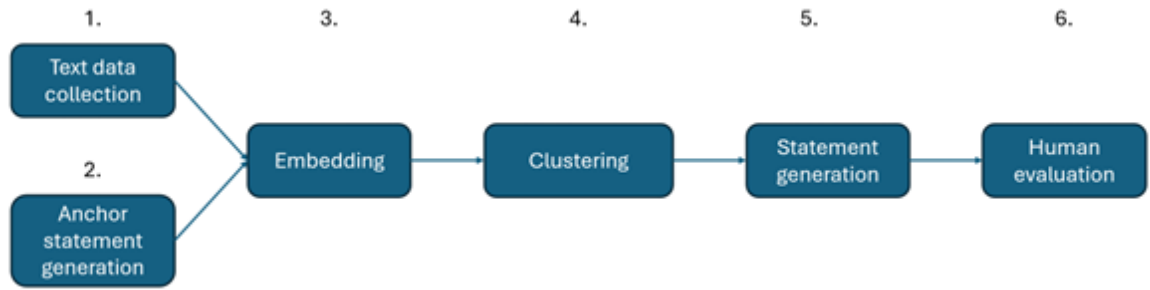

*Figure 2: The 6 steps involved in order to generate and validate DRI statements automatically*

The following sections of this chapter provide an in-depth description of each stage of the process, detailing the data collection and preprocessing methods, computational techniques, and evaluation procedures employed in the study[4].

**3.2 Data Collection**

### 3.2.1 Corpus Selection & Compilation

The selection of an appropriate data source is crucial for the accurate capturing of the public discourse on the given topic. For this study, news articles were chosen as the primary data source for several reasons. News articles suit the needs of the project particularly well since they already come in written form and are therefore closer in form to the intended output as say TV or Radio shows. This alignment facilitates a seamless transition from data collection to analysis and to the generation of DRI statements and policy proposals. Additionally, news outlets offer a wide range of public opinions and arguments, reflecting the diversity of societal perspectives on a given topic. They serve as a rich repository of information, encompassing viewpoints from different demographic groups, regions, and political affiliations. This diversity is essential for creating a comprehensive and balanced DRI survey that accurately represents the multifaceted nature of the public debate on the given topic. Moreover, the quality of journalistic writing ensures that the data is coherent, reliable, and of high informational value. Professional journalists usually adhere to standards of accuracy and objectivity, which enhances the reliability of the content extracted for analysis. This is especially true, since we decided to only consider articles from well-established news outlets such as NZZ, SRF and WOZ. Lastly, news articles are more generalizable than qualitative interviews with individual participants. While interviews can provide in-depth insights, they may not capture the broader public sentiment, or the variety of opinions present in public discourse. It is acknowledged that some unique or minority viewpoints might be underrepresented in news outlets, which could turn out to be a limitation of this approach (Buyens & Aelst, 2021). Considering that the topic at hand was closely related to Switzerland, we gathered our text corpus from the Swissdox@LiRI database, which provides easy accessibility to a vast database of Swiss news articles. We restricted the news outlet selection to a timeframe spanning from January 1, 2018, to August 29, 2024, capturing only recent developments and contemporary discussions surrounding the given topic of health care costs. A set of seventeen keywords related to various aspects of the debate on health care costs was employed to query the database in order to only collect articles that include at least one of the keywords. Keywords included several synonyms for words like 'hospital', 'health insurance' and 'health politics'. To ensure all language regions of Switzerland are represented in the dataset, we included each keyword in German, French and Italian[5]. This keyword search resulted in a collection of 444'869 unique news articles.

---

[4] The source code is available on GitHub: https://github.com/mooorice/automated-DRI-statement-generation

### 3.2.2 Data Preprocessing & Embedding

After assembling the corpus of news articles, the next step of the method involved transforming the textual data into a format suitable for computational analysis. To achieve this goal, the text data was first chunked into paragraphs and then embedded using an embeddings model.

The initial phase of data preprocessing includes segmenting the articles into individual paragraphs. The decision to chunk into paragraphs and not e.g., sentences, is based on the understanding that paragraphs represent more meaningful aggregations of information compared to sentences. A paragraph often encapsulates a complete idea or argument, providing a coherent unit of thought valuable for semantic analysis. By focusing on whole paragraphs, a language model can concentrate better on the meaning of the text rather than the syntactic structures, specific vocabulary or stylistic nuances present at the sentence level. This allows for the extraction of key semantic information which is essential for generating meaningful DRI statements and policy proposals. After chunking the articles into paragraphs, another keyword search selects all paragraphs that include at least one of the keywords from the list to ensure relevancy of each paragraph to the topic at hand. The process of chunking the dataset of news articles resulted in a substantial collection of 738 '598 unique paragraphs with a median of 1 paragraph per article and a mean of about 2.86.

---

[5] The list of keywords is available in Appendix A.

To enable computational manipulation and analysis, the textual paragraphs need to be transformed into numerical representations. This transformation is achieved through a process known as embedding, where each paragraph is converted into a high-dimensional vector. These embeddings capture the semantic essence of the text, allowing the computational model to interpret and compare paragraphs based on their content (Wang & Koopman, 2017). Specifically, each paragraph is mapped to a 384-dimensional vector space using the [paraphrase-multilingual-MiniLM-L12-v2 model](#). This model is a compact and powerful transformer-based language model from the SBERT-family (Sentence-BERT) (Reimers & Gurevych, 2019) designed for generating high-quality sentence and paragraph embeddings across multiple languages. Several reasons motivated the selection of this particular model. Firstly, given that the corpus includes articles in German, French, and Italian, a multilingual model was essential. The paraphrase-multilingual-MiniLM-L12-v2 model is trained on a diverse set of languages, ensuring it can handle the multilingual nature of the dataset without compromising the quality of embeddings. Secondly, the model's general-purpose utility makes it suitable for a wide range of natural language processing tasks, including semantic similarity analysis, clustering and information retrieval, aligning well with the objectives of this study. Furthermore, despite its robust performance, the model has relatively low computational requirements. This efficiency is crucial when processing a large corpus consisting of hundreds of thousands of paragraphs, as it reduces the computational resources and time needed for embedding. Additionally, recent comparison of word embeddings models has shown that, for semantic analyses, SBERT is still one of the most capable models, outperforming many LLMs (Mahajan et al., 2025). Lastly, the model enjoys a large user base and is popular on platforms such as Hugging Face, a leading repository for machine learning models. This popularity ensures that it is well-documented and supported by a community of developers and researchers, facilitating implementation and troubleshooting.

## 3.3 Dimensionality Reduction & Categorization

### 3.3.1 Dimensionality Reduction

The embeddings generated by the model comprise 384 dimensions. While these high-dimensional embeddings capture intricate semantic details of the textual data, they present significant challenges for computational analysis. High-dimensional data often introduces noise and redundancy, which can obscure meaningful patterns and relationships within the dataset (Shiebler et al., 2018). Additionally, such data can be computationally intensive to process, requiring substantial memory and processing power. To mitigate these challenges, we reduce the dimensionality of the embedding space to fifty dimensions. Dimensionality reduction serves to simplify the data, decrease computational load, reduce noise and enhance the clarity of underlying patterns, thereby improving the efficiency and effectiveness of the analysis.

In this study, the Uniform Manifold Approximation and Projection (UMAP) algorithm was employed for dimensionality reduction. UMAP is a non-linear technique that excels at preserving both the global and local structures of high-dimensional data while reducing its complexity (McInnes et al., 2020). This reduction not only decreases computational demands but also enhances the visibility of patterns within the data. UMAP was chosen over other dimensionality reduction methods due to its ability to maintain the intrinsic topological structure of the data. Unlike linear techniques such as Principal Component Analysis (PCA), UMAP can capture non-linear relationships inherent in semantic embeddings of natural language data (ibid.). This capability is crucial for preserving the semantic relationships between paragraphs, ensuring that the reduced-dimensionality data remains representative of the original high-dimensional embeddings.

Several parameters needed to be selected to optimize UMAP's performance for this specific application. The number of components was set to 50, striking a balance between retaining sufficient complexity to preserve meaningful semantic relationships and reducing the data's dimensionality to a level that mitigates noise and computational overhead. The 'number of neighbors' parameter was set to 30. This parameter determines the balance between local and global structure preservation in the data. After experimenting with different values, a value of 30 was found to be an effective middle ground, providing a reasonable compromise that preserves local neighborhood structures while still capturing broader global patterns. The minimum distance parameter was set to 0.0, indicating that UMAP should preserve as much of the data's topological structure as possible. This setting is particularly important when the goal is to find clusters, as it allows points to be placed close together in the reduced-dimensional space if they are similar in the high-dimensional space. Preserving fine topological structures ensures that the nuances of semantic relationships are maintained. Cosine similarity was employed as the distance metric in UMAP. Cosine similarity measures the cosine of the angle between two vectors, focusing on their orientation rather than their length. This choice was made because cosine similarity is more effective than Euclidean distance in sparsely populated spaces, especially when dealing with text embeddings where the direction of the vector is more informative than its magnitude (Ismael, 2025). This metric also aligns with the similarity measures used later in the analysis when calculating the semantic similarity between paragraphs and anchor definitions during the categorization process.

### 3.3.2 Categorization by Cosine Similarity

After reducing the dimensionality of the paragraph embeddings, we categorize the paragraphs into relevant subtopics within the broader theme of the given topic. The goal here is to group similar paragraphs together based on their semantic similarity to the anchor statements, ensuring that each category represents a specific sub-discourse of the debate.

The categorization step is fundamental to our methodology as it directly impacts the diversity, quality and relevance of the DRI statements and policy proposals generated later in the process. It also allows for the organization of the vast corpus into manageable and thematically coherent groups. Given the extensive size of the dataset, with hundreds of thousands of paragraphs, an unsupervised clustering approach without guidance can easily lead to incoherent or irrelevant groupings due to the complexity of the data. The learnings from experimentation with such an approach can be found in chapter 3.6. Additionally, the categorization step facilitates the extraction of meaningful insights by isolating specific aspects within the broader discourse on the given topic. This is crucial for generating accurate and representative DRI statements according to the predefined subcategories of the survey.

To achieve effective categorization, our method employs a supervised approach using anchor definitions, drawing on the work by Aceves and Evans (2024). These anchors serve as semantic reference points that guide the categorization process to make sure that paragraphs are grouped based on their relevance to predefined subcategories of the health care cost debate. The first step in this categorization methodology is the creation of anchor definitions for each of the predefined subcategories. In our case these amount to seven specific themes within the topic of health care costs and one overarching category on health costs in general. These subcategories were identified based on their significance in public discourse and policy debates. The anchor definitions were generated using an LLM (Large Language Model), specifically prompted to produce summarizing paragraphs for each subcategory.

The LLM was provided with detailed instructions and context to generate accurate and comprehensive anchors. For each subcategory, the model received the heading of the category and a 50-page policy field analysis paper produced by 'Interface,' an independent competence center for evaluation in Switzerland (Bischof et al., 2024). This paper was specifically prepared for the organizers of the 'Bevölkerungsrat 2025'. The use of this policy field analysis ensured that the anchor definitions were grounded in expert analysis and reflected the key issues and arguments from each subcategory. The LLM then generated summarizing paragraphs in German, French, and Italian for each subcategory, aligning with the multilingual nature of the corpus. These anchors encapsulated the central themes, concerns, and debates associated with each subtopic, providing a clear semantic representation of what each category should encompass. Once generated, the anchor definitions were added to the dataset of news article paragraphs to undergo the same embedding and dimensionality reduction process as our text corpus[6].

With both the paragraph embeddings and the anchor embeddings prepared, the next step involves calculating the cosine similarity between each paragraph embedding and each of the eight anchor embeddings. For each paragraph, the cosine similarity to each anchor was computed, resulting in eight similarity scores per paragraph, i.e. one for each subcategory. These scores quantified how closely the content of each paragraph aligned with the themes represented by the anchors. This categorization approach provides several advantages. It leverages the strengths of supervised and unsupervised methods by guiding the process with predefined semantic anchors while allowing the data to organize itself based on inherent similarities. Additionally, the use of anchor definitions ensures that the categories are meaningful and aligned with the study's objectives, reducing the risk of forming arbitrary or irrelevant groupings. Moreover, calculating cosine similarity is computationally less expansive than more complex clustering algorithms. Also, the way we approached the categorization allows for paragraphs to score high in more than one category, ensuring that paragraphs with a high relevance on several subtopics will be counted in each of those categories[7].

---

[6] The prompts and anchor statements are available in Appendix B.

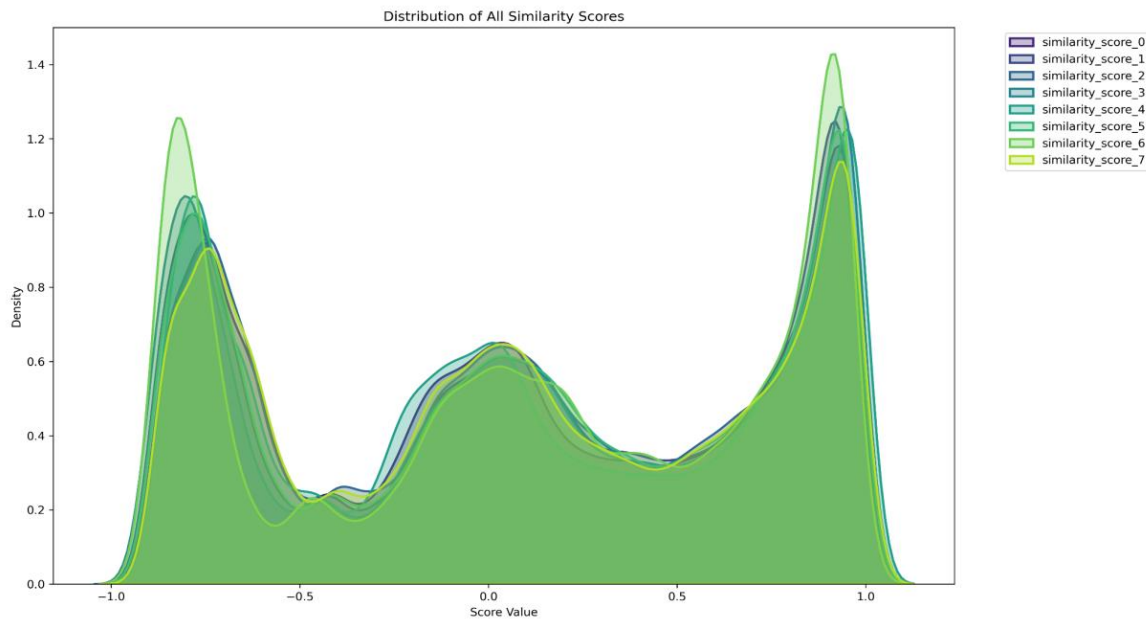

*Figure 3: Distribution of similarity scores across the dataset. High similarity scores indicate high similarity to the anchor statement indicating high likelihood of a relevant input to the discourse on the defined topic.*

### 3.4 Automatic Generation of DRI Statements & Policy Proposals

After categorizing the paragraphs according to their semantic similarity to the anchor definitions, the next step involves the automatic generation of DRI statements and policy proposals. Here our method transforms the categorized textual data into statements and policy proposals that reflect the public discourse on the given topic. The objective is to generate concise, representative statements that encapsulate the core arguments and policy suggestions within each subcategory and political leaning of the media landscape.

To ensure that only the most relevant and representative content is passed to the LLM for statement generation, we selected paragraphs with the highest cosine similarity scores for each of the eight categories. While the top 50'000 paragraphs all showed strong similarity (scores above 0.92), they had approximately 80% overlap between categories. This high overlap meant that using a 50,000-paragraph threshold would result in the LLM receiving largely redundant information across different category prompts, possibly leading to the generation of very similar DRI statements across the different categories. To pass only paragraphs uniquely fitted to the relevant category, we significantly raised the selection criteria to include only the top 500 highest-scoring paragraphs per category. This stricter threshold reduced the average overlap between categories to approximately 14.5%. The comparative overlap patterns at different thresholds are visualized in the heatmaps presented in Figure 4.

---

[7] See Figure 3 for a visualization of the distribution of similarity scores for each category.

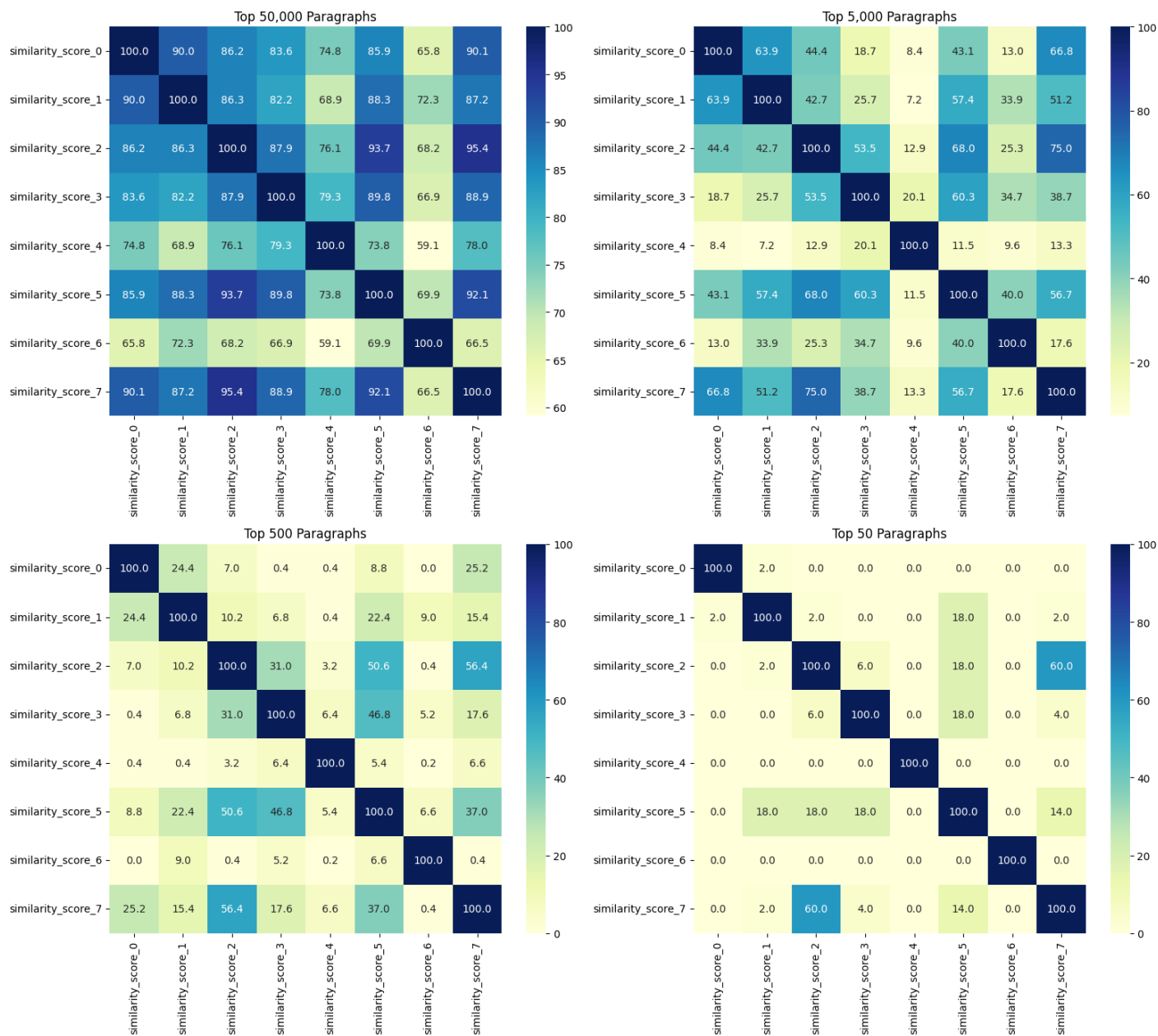

*Figure 4: Overlap of paragraphs in the top x paragraphs per similarity score.*

As the amount of coverage and available text data can vary greatly between the different political leanings of news outlets, the division of our data into several political camps is essential to ensure equal consideration of different political standpoints. We divided our dataset by political leaning of the given media outlet into the five categories left, left-liberal, centrist, right-liberal and right. To achieve this division, we relied on a recent study by Udris (2023) comparing the 24 of the most influential news outlets of Switzerland in terms of political leaning on Swiss referenda. The continuous scale of the left-right dimension reaching from -100 to 100 was coded into 5 categories, categorizing each newspaper outlet according to its scoring with the centrist category lying between (-5, 5), the right- and left-liberal categories falling between [5, 15], respectively [-5,-15] and all news outlets scoring higher than 15 or lower than -15 as right respectively left. We only considered paragraphs that appeared in media outlets analyzed in the study.

To generate the DRI statements and policy proposals, two specialized assistants were created building on the GPT-4o large language model (Hurst et al., 2024). An assistant is an entity within OpenAI's API landscape that comes with several additional key functionalities compared to the standard chat_completions endpoint. One of which is the option to attach files to a prompt, which is particularly important to our use case.

The first assistant, referred to as the 'Considerations Assistant' is tasked with producing DRI consideration statements. The system prompt provided to this assistant includes a comprehensive explanation of the DRI concept, emphasizing the importance of capturing the reasoning and deliberative aspects of the public discourse. The prompt defines the desired form of the consideration statements, highlighting key characteristics such as neutrality, conciseness, and representativeness. Examples of well-crafted DRI statements were included alongside bad examples within the prompt to guide the assistant, ensuring consistency in style and content. For additional context, we attached a collection of about 650 example statements from previous DRI surveys, stored in a vector database. The task is to produce five consideration statements in JSON (JavaScript Object Notation) format, which is a data format that stores and transmits structured information using key-value pairs in a human-readable and machine-parsable text format.

The second assistant, called 'Policy Assistant', is responsible for generating policy proposals derived from the categorized paragraphs. Like the Considerations Assistant, the system prompt for the Policy Assistant elaborates on the DRI concept, particularly focusing on the goal of formulating actionable and practical policy options. The prompt outlines the expected format of the policy proposals, stressing the need for clarity, feasibility, and alignment with the public's concerns and suggestions. A list of about 100 policy proposals from previous DRI surveys, stored in a vector database, was attached to provide context as to how policy proposals should be formulated. The assistant is then instructed to produce five policy options in JSON format.

An important consideration in this step is the potential for inherent biases within the language model, which could potentially influence the framing of statements if not properly managed. To mitigate this risk, the political leaning of the target output was explicitly specified in the prompts provided to the assistants. By clearly indicating the desired perspective, the assistants are guided to generate statements that accurately reflect the viewpoints associated with each political orientation. To this end and to streamline the generation process in general, a flexible prompting function was developed. This function prompts the assistants to produce outputs for each subcategory and political leaning in a single run, optimizing efficiency. The function dynamically adjusts the prompts based on subcategory and political leaning, ensuring that the assistants receive accurate and context-specific instructions for each task.

We selected GPT-4o as the language model of our choice due to its status as one of the most powerful and sophisticated models available at the time of the study. GPT-4o's advanced capabilities in language comprehension and generation, coupled with its proficiency in handling multilingual data and its wide context window, make it exceptionally well-suited for this task (ibid.). The model's ability to understand and generate text in German, French, and Italian ensures that the assistants could effectively process the multilingual dataset and produce coherent and contextually appropriate statements. Moreover, the availability of advanced API connections and the Assistants API facilitated the development of complex setups involving multiple assistants, system prompts and vector databases. These features allow for a high degree of customization and control over the assistants' behavior, enabling the researchers to control the generation process in order to meet the specific requirements of the study.

During the generation process, each assistant processes the selected top 500 paragraphs for each subcategory and political leaning, applying their respective instructions to produce the desired outputs. The Considerations Assistant generates five DRI consideration statements per set, while the Policy Assistant produces five policy proposals. This process results in a comprehensive collection of statements that spans the full range of topics and political perspectives encompassed in the study.

**3.5 Final Selection**

After the DRI statements and policy proposals are generated, they are handed to a group of experts for human evaluation and final selection. This step is crucial to ensure accuracy, relevance, and quality of the outputs. The generation process resulted in a list of 200 DRI statements and 25 policy options, encompassing a general category and seven subcategories, each articulated from five political perspectives. Since some key narratives appear in several news outlets across the political spectrum, it was expected that some of the statements are closely related. To mitigate the issues of quality and repetition of narratives, human evaluation is essential to ensure a high quality DRI survey. To this end, the evaluation team reviewed each statement. The team selected the statements that most effectively captured the essence of the issue while keeping a balance of ideology and categories in the survey. The same approach was applied to the selection of policy options, aiming to include the most prominent ideologies in the final selection of 8 policy options. This process reduced the high number of statements to 28, which is more suitable for DRI surveys. By refining the collection in this manner, the study ensured that each statement added unique value to the DRI survey.

While automation can process large volumes of data and generate initial outputs at an unprecedented scale, human judgment is essential to ensure the quality and applicability of these results. The systematic approach taken by this method exemplifies best practices in integrating computational methods with scholarly scrutiny. The methodology also shows how technology and human expertise can synergistically produce outcomes that are both comprehensive and nuanced, offering a model for future studies seeking to analyze complex public discourses with the help of LLMs.

**3.6 Dead Ends & Learnings**

While developing the methodological framework for this study, several alternative approaches were explored to further analyze the embedding space generated from the corpus of news article paragraphs. Notably, attempts were made using clustering algorithms and topic modeling techniques such as HDBSCAN (Hierarchical Density-Based Spatial Clustering of Applications with Noise) (McInnes et al., 2017) and BERTopic (Bidirectional Encoder Representations from Transformers - Topic) (Grootendorst, 2022). While these methods showed initial promise, they ultimately did not yield relevant results, providing valuable learning experiences that informed the final methodology.

Initially, clustering was considered a viable approach to make sense of the high-dimensional embedding space. The goal was to identify distinct clusters within the data that corresponded to specific discourses or topics present in the corpus. By uncovering these clusters, we hoped that the analysis could reveal underlying thematic structures and facilitate the generation of more targeted DRI statements. The clustering algorithm selected for this purpose was HDBSCAN (McInnes et al., 2017). HDBSCAN is an extension of the DBSCAN algorithm and well-suited for clustering large datasets with complex structures. It operates by identifying areas of high density in the data and grouping points accordingly. This can be advantageous when dealing with non-linear relationships and varying cluster densities. Upon applying HDBSCAN to the dimensionality-reduced embeddings, the results were not as informative as anticipated. The clusters identified tended to revolve around specific keywords or named entities rather than coherent topics or arguments related to health care policy. For example, some clusters were dominated by mentions of pharmaceutical company names, government ministries, or other proper nouns. This pattern suggested that the clustering algorithm was capturing surface-level lexical similarities rather than deeper semantic or thematic connections.

Further analysis using centroid examination supported this observation. The centroid statements, i.e. the central and therefore in theroy most representational points of each cluster, did not exhibit particularly meaningful or distinctive characteristics that could explain or define the clusters in terms of substantive content. Instead, they often reflected the same keyword-driven patterns observed in the clustering results. This outcome supported the conclusion that the clusters were not aligned with the complex, topic-level distinctions necessary for generating insightful DRI statements.

These findings were further reinforced by the application of BERTopic (Grootendorst, 2022), a topic modeling technique that integrates clustering with natural language processing. In addition to the aforementioned clustering, BERTopic incorporates keyword extraction for each cluster, aiming to identify and represent topics within the data effectively. The motivation for considering BERTopic stemmed from its design to handle large textual datasets and its potential to uncover latent topics without extensive manual intervention. By leveraging advanced embedding models and clustering algorithms, BERTopic promises a more nuanced understanding of the data, which could have enhanced the thematic organization and interpretation of the corpus. However, upon applying BERTopic to the dataset, similar issues emerged as those encountered with the initial clustering approach. Since BERTopic was analyzing the clustering output by HDBSCAN, the clusters identified by BERTopic were again predominantly driven by keywords or named entities, rather than cohesive topics or arguments relevant to health care costs. The keyword extraction process highlighted terms that, while common within the data, did not necessarily coalesce into meaningful themes suitable for the study's objectives. The reliance on surface-level lexical features resulted in clusters that lacked the depth and specificity needed to inform the generation of DRI statements and policy proposals. The clusters did not align well with the substantive issues explored in the health care cost debate, limiting the utility of BERTopic for this particular application.

Comparing the clustering results to the categorization approach with anchor statements, we found that the maximum difference between any two categories' similarity scores for any of the clusters was only 0.185 within the top 20 clusters with the highest spread having relatively low absolute similarity values ($\leq 0.6$) across all categories. Furthermore, looking for clusters with high similarity scores (0.9) for one category while constraining all others to lie below a smaller similarity score yielded only two items with a relatively high threshold of 0.85 for lower scores. These findings strongly suggest that unsupervised clustering can't produce clusters that correlate well with our predefined categories. This is not to say that BERTopic with HDBSCAN is generally ineffective. However, if one already knows what the data should be clustered on (i.e., has specific categories in mind), an unsupervised method may not be the most suitable approach. In contrast, calculating cosine similarity to predefined category embeddings directly addresses the question of fit to those categories, making it a more targeted technique for our particular use case.

The experiences with clustering, centroid analysis, and BERTopic provided valuable insights into the challenges of analyzing large, high-dimensional textual datasets. One key learning was the recognition that unsupervised clustering algorithms, even advanced ones like HDBSCAN, may struggle to capture the complex semantic relationships inherent in natural language data, especially when the goal is to identify predefined topics. These methods tended to group data based on superficial lexical similarities, such as shared keywords or named entities, rather than deeper thematic connections. Another important realization was the significance of guided approaches in analyzing textual data for very specific research objectives like ours. The clustering attempts lacked the incorporation of domain knowledge or predefined categories, which turned out to be crucial for aligning the analysis with the study's goals. In contrast, the successful use of anchor definitions and cosine similarity in the final methodology provided a more directed and effective means of organizing the data according to relevant themes.

## 4. Discussion

### 4.1 Evaluation

Apart from the human evaluation during the final selection of statements, we ran our method on a second topic to compare the outputs to an existing DRI Survey to further validate our method. The selected DRI survey was conducted alongside the Swiss Citizens' Assembly for Food Policy (SCAFP). The Citizens' Assembly took place over a six-month period in 2022 and focused on 'The Future of the Food System in Switzerland.' The questions used in the SCAFP were developed by the author of the original study Philippe Mathys and were based on a reconstruction of the public debate surrounding the Swiss food system (Mathys, 2024). This reconstruction drew on general themes of trust, responsibility, justice, and food systems, as well as central themes identified in recent popular initiatives in Switzerland that were documented in official voting booklets (ibid.). In addition, the study incorporated highly supported statements from opinion surveys conducted by GFS.BERN and Tamedia (ibid.). The DRI survey's structure ranged from very general statements on climate change to more specific positions on food policy, mirroring the layered nature of public discourse on this issue (ibid).

To search for relevant news articles in the Swissdox API, we first designed a prompt to generate a list of fitting keywords in German, French, and Italian. The articles were then split into paragraphs. After filtering out duplicates and paragraphs that did not include any of the keywords, we were left with a dataset of 393'258 paragraphs. For anchor generation we collected all handouts from the Citizens' Assembly's scientific inputs and combined the dataset with chapter 3 from Mathys' paper which provided additional context on the issue at hand. The existing sub-headings from Mathys' DRI Survey served as categories for anchor generation. The approach was deliberately left rudimentary in order to leave room for shortcomings of the method through the prevention of overspecification by the applicant[8].

Once the textual corpus and anchors were established, each paragraph was embedded using the same model as above. The Dimensionality was reduced to 50 using UMAP and afterwards cosine similarity was calculated between each paragraph and each anchor. The process of generating new DRI statements proceeded through the OpenAI Assistants API. Due to new constraints on the usage of the GPT4o-model, we employed the smaller GPT4o-mini model instead. By picking a smaller, less performant, model as an alternative, we made sure that the validation results would, if at all, only be biased towards worse outcomes. Minimal changes were made to the prompts in comparison to the health-care-costs case study, with the category names and the topic ('Future of Switzerland's Food System') being the main elements adapted to the new topic at hand.

The generated statements were subsequently evaluated against the original statements from the SCAFP DRI Survey. All statements from both the newly created pipeline and the pre-existing DRI Survey were again embedded using the paraphrase-multilingual-MiniLM-L12-v2 model. Cosine similarity scores were calculated for every possible pairing of original and newly generated statements. A manual review then examined the five closest generated statements for each of the original DRI items. Each statement was compared manually, focusing on how closely the generated statements matched the intent, content, or nuance of the original. Out of 41 original DRI statements, the pipeline yielded 14 that were considered good fits, 3 that were good-to-okay fits, and 3 that were okay fits, amounting to 20 matches in total (approximately 50% of the original set). Of the 8 policy preferences in the original survey, 5 aligned well and 2 were deemed acceptable matches, leading to a total of 7 matches (87.5%)[9]. Considering that the generation of DRI statements is a highly subjective task that leaves a lot of room for different interpretations, the fact that we employed a weaker model for the generation and did not offer any further information than the mostly unspecific titles of the subcategories, we deem the results as highly impressive.

A closer examination of these results revealed that 12 of the 41 original statements and 1 of the 8 policy preferences were formulated in a very general style that was not specified in the prompts given to GPT. The LLM aimed at specific formulations to remain close to the wording and structure of the pipeline's core method. Here the headings of the different sections of the DRI survey did not pose enough context for the model to adjust the topic and the style of DRI statements. By excluding these highly general items, the remaining subset included 29 statements and 7 preferences, from which the pipeline successfully matched 19 statements (about 65%) and all 7 preferences (100%). Especially where there were concrete topics touched by a statement, for example in Q30_3 "Agricultural subsidies should promote sustainable, pesticide and antibiotic-free production." The method performed exceptionally well with several generated statements expressing closely related variations of the statement such as "Investments in ecological farming methods should be prioritized over subsidies for conventional agriculture.".

---

[8] The keywords and anchor statements are available in Appendix C.

[9] The source code is available on GitHub: https://github.com/mooorice/automated-DRI-statement-generation/tree/main/Validation/data/06_comparison

Analyzing the mismatches, we can tell that our method was lacking instructions to produce highly generalized statements and to craft personal opinion statements from a first-person perspective. It also lacked instructions to generate pessimistic or negative formulations, such as assertions about the incompatibility of a sustainable food system with accessible food pricing and security. These shortcomings point to the model's tendency to default to positive or neutral framings and a lack of diversity in formulation. These shortcomings could possibly be mitigated with more specific prompting of the models. It is, however, not necessarily required for DRI surveys to include these diverse forms of framing of statements. In addition, alternative or more fringe narratives, for instance those positing that global warming might benefit food production, failed to emerge in the generated statements likely because such perspectives were underrepresented or altogether absent in the source media data. Another reason for these shortcomings might be the inherent bias of the GPT4o model, leading to neutral or possibly one-sided statements.

Another circumstance led to further indication that the proposed method produces valuable representations of the discourse. Due to constraints on the mid-term DRI survey length for the 'Bevölkerungsrat 2025', the amount of questions had to be minimized. The goal was to shorten the number of questions while capturing as much variance as possible. The approach was to minimize the amount of questions from the first DRI survey of which the results were already available, while maintaining consistency with these results. In the process, which is described in more detail by Veri (2025), a multifactor analysis was employed to identify underlying dimensions, resulting in eight distinct factors with eigenvalues above one. These results can be understood as a further indication that the method produced DRI statements that capture at least eight distinct dimensions of the discourse on the given topic.

Taken together, these findings underscore both the potential and the limitations of using AI-driven pipelines for generating survey statements in the context of deliberative reasoning indexes. The good matches between AI- and human-generated DRI statements highlight the model's capacity to identify and reformulate core arguments from a large corpus of text data, especially when provided with enough concrete context. Additionally, the large amount of distinct dimensions captured by the DRI statements indicates that the proposed method can capture a wide range of aspects on a given topic. On the other hand, the absence of certain negatively or highly generally formulated statements reveals the importance of human quality control, carefully crafted prompts and potentially more diverse training data. Particularly niche or radical viewpoints may remain elusive if they are not present in mainstream media sources, a point that carries important implications for the method's comprehensiveness and inclusivity in capturing the full range of public discourse. The deviations in the style of formulation can however at least partially be accredited to different approaches to the formulation of DRI statements that do not necessarily pose an issue in terms of quality of any of the two outputs compared.

## 4.2 Limitations

While the methodology developed in this study offers significant advancements in automating the generation of DRI statements and policy proposals, it is important to acknowledge several limitations that may affect the generalizability and applicability of the findings. A primary constraint of the method is its reliance on access to a large, high-quality text dataset on the specific topic under investigation. In this case, the availability of an extensive corpus of news articles on health care costs in Switzerland was instrumental in achieving meaningful results. However, such comprehensive and readily accessible datasets may not be available for all topics or regions. Especially if the topic at hand is less prominent. In situations where the luxury of a free or affordable large-scale database is absent, the cost and effort required to acquire sufficient data can be prohibitive. This limitation restricts the method's applicability to well-documented subjects and may hinder its use in research areas where data scarcity is an issue. Relying on news articles as the basis for analysis may therefore prove particularly unreliable in countries with a more biased or less developed media landscape. Considering the ever-growing role of social media for public discourse, news outlets might further lose their significance and therefore their representativeness in the long run, even in countries where they play an important role as of now.

Moreover, the use of news articles as the primary data source, while advantageous in capturing a wide range of public discourse, may not provide the depth of insight that can be obtained through in-depth qualitative interviews. News content often presents information at a surface level due to constraints such as word limits and editorial policies. As a result, the generated DRI statements and policy proposals might lack the nuanced understanding of individual perspectives and the underlying motivations that qualitative interviews can reveal. This superficiality could lead to an incomplete representation of the complexities inherent in public reasoning on the topic.

Another limitation concerns the potential underrepresentation of certain opinions within mainstream media outlets. News outlets, by nature, may not fully capture the voices of minority groups or those holding niche or less popular viewpoints. Consequently, the methodology might overlook important perspectives that are not prominently featured in traditional media. This gap can affect the inclusivity and comprehensiveness of the analysis, potentially skewing the findings toward more widely held or dominant opinions. This is an issue that has emerged during the validation process as well. Here we noticed that statements referring to more fringe perspectives, such as Q60_6 "Climate warming will ultimately benefit food production globally as plants better grow in higher temperature", did not appear in the AI-generated statements.

To address these limitations, it remains an open question whether incorporating lower-quality data sources, such as social media comments, could enhance the methodology. Social media platforms often host a diverse array of opinions, including those from underrepresented or marginalized groups. Integrating such data could help capture a broader spectrum of viewpoints and contribute to a more inclusive set of DRI statements and policy proposals. However, this approach introduces challenges related to data quality, as social media content can be noisy, unstructured, and variable in reliability. Especially with developments like Meta distancing itself from fact-checking (Isaac & Schleifer, 2025) and political camps moving to distinct services, it seems like the quality of such data might further decline. Ensuring that the inclusion of such data leads to high-quality results would require the development of robust filtering and validation techniques to mitigate potential issues.

In summary, while the methodology demonstrates considerable potential, its dependence on extensive, high-quality textual data limits its generalizability to topics with abundant media coverage and accessible datasets. The possible superficiality of the extracted information and the underrepresentation of certain opinions in mainstream media could further constrain the depth and inclusiveness of the findings. Future research could explore strategies to overcome these limitations, such as incorporating alternative data sources and enhancing data processing techniques, to broaden the applicability of the method and improve the richness of the insights generated.

## 4.3 Implications

Regardless of their constraints, our findings have profound implications for the practice of measuring the outcome of deliberative processes. By introducing a method that automates the generation of DRI statements and policy proposals the research helps to simplify the preparation of DRI surveys. This automation reduces the time and resources traditionally required, making it feasible to apply DRI surveys more widely across various deliberative contexts.

The ease of preparing DRI surveys through this method could result in a much broader application of DRI as an evaluative tool. The method lays a foundation for a standardized process for the preparation of DRI surveys which lowers the barriers to deployment and improves comparability between studies. Therefore, researchers and practitioners can implement DRI surveys in diverse settings, enabling comparative studies that were previously impractical due to logistical constraints. Such comparative analyses are invaluable for deepening our understanding of deliberative processes, particularly regarding how design decisions and contextual parameters impact DRI outcomes. By facilitating these studies, the method contributes to the refinement of deliberative practices and supports the development of more effective deliberative forums.

Moreover, the integration of ML and AI methods into the assessment of deliberative processes opens new avenues for enhancing digital deliberation platforms. The method allows for the provision of high-quality feedback data to computational models, enabling them to be trained on DRI outcome quality. By automatically incorporating data on the impact of their actions, these models can iteratively improve their performance in fostering higher-quality deliberation outcomes. This capability could potentially improve digital deliberation tools by making them more responsive and adaptive to the dynamics of participant interactions, ultimately enhancing the quality of public discourse.

Another significant implication of the method is its contribution to the representativeness and comprehensiveness of the DRI survey. By systematically processing all available media outputs relevant to the topic, the method ensures that a wide array of reasons and perspectives are included in the survey instrument. This comprehensive coverage provides a strong basis for asserting that the DRI survey has not overlooked any major facets of the discourse, aside from the limitations previously discussed. As a result, the DRI measure becomes more self-contained, meaning it fully encapsulates the relevant considerations within the public sphere.

A self-contained DRI measure enhances the accuracy and stability of the evaluation. When the survey thoroughly represents discourse, any misalignment of policy preferences among participants can be more directly attributed to differences in their consideration statements and vice versa. This clarity strengthens the validity of the DRI as an outcome measure, providing more reliable insights into the effectiveness of deliberative processes. It allows researchers to draw more precise conclusions about how and why participants' views evolve during deliberation, informing strategies to improve engagement and reasoning in future initiatives.

Beyond its immediate applications in deliberative democracy research, the method has broader implications for political science, particularly in the field of text analysis and computational social science. Our approach enables the systematic and scalable processing of large corpora of textual data, making it a valuable tool for studying political discourse, public opinion formation, and media influence. Through the definition of anchor statements, the method provides a structured way to navigate vast and complex political discussions. This does not only improve the precision of content analysis but could also allow for the dynamic tracking of shifting narratives over time. As political science increasingly engages with big data and AI-driven methodologies, such approaches offer powerful means to uncover latent structures in discourse, assess ideological framing, and enhance our understanding of the mechanisms shaping political communication and decision-making processes.

The method developed in this study therefore not only advances the practical application of the DRI but also enriches the toolbox of computational social science in general. It demonstrates the feasibility and benefits of integrating advanced computational techniques into the evaluation of deliberative democracy. By making the assessment process for DRI more efficient, comprehensive, and adaptable, the method paves the way for more robust and insightful analyses of deliberative practices. It encourages the continued exploration and adoption of ML and AI in this field, with the promise of enhancing both the quality of deliberation and our understanding of its underlying mechanisms.

## 4.4 Outlook

Building upon the findings and limitations of this study, several avenues for future research emerge that could enhance the understanding and application of the DRI in evaluating deliberative processes. First of all, the method itself, even though it's already producing impressive results, was severely limited by performance- and time constraints during development which leaves some room for improvement. Some research suggests that different dimensionality reduction algorithms and higher dimensionalities could potentially improve the quality of the semantic similarity search results (Huertas-García et al., 2023). Future applicants of the method should therefore put some more research into finding the most suitable algorithm and parameters. Additionally, limiting the amount of news paragraphs provided per request to 500 seems rather low given that modern LLMs are able to handle much larger datasets. For future applications it might be useful to improve our understanding of the trade-offs between the overlap of categories and size of the dataset. Another place for improvement could be the application of a more observable and steerable LLM for the generation of DRI statements. With OpenAI's closed models it is impossible to understand when and how the model accesses the provided data and how much of it it uses for generation. The application of a more transparent model could help to further improve the representativeness of the outputs while also helping researchers in understanding the decision making of the model. Considering the probabilistic nature of LLMs, it's highly likely that the outputs of the model will vary when prompted several times. Therefore, repeating the generation process several times and sampling the most frequently appearing DRI statements could further improve the results of the method.

Regarding the growing impact of social media, it might be highly beneficial for our method to investigate the impact of incorporating lower-quality data sources, such as social media comments, into the DRI survey generation process. This approach could make the method much more versatile and representative and therefore improve its impact significantly. Social media platforms contain a wealth of diverse opinions on virtually any possible topic. However, the data is often unstructured and varies in quality. Future studies could examine methods for filtering and refining this data to maintain the integrity of the DRI measure. Exploring this avenue could make the application of the DRI more cost-effective by reducing reliance on expensive, high-quality datasets. By finding ways to include underrepresented or niche opinions captured in social media, researchers might also enhance the inclusivity and representativeness of the DRI surveys without compromising on quality. Thinking further ahead, an integration of social media data within the method could become a first step in analyzing social media discourses themselves through the lens of DRI.

Another promising direction is the comparative analysis of participants' survey responses with their actual contributions during the deliberation sessions. This approach would require the transcription and analysis of the entire deliberative process, enabling the identification of high-impact moments, statements, and participants. Such an approach would be particularly easy to pursue in the analysis of social media debates since discussion is already written down. By correlating these elements with participants' answers in the DRI survey and its outcomes, researchers could gain insights into the specific aspects of deliberation that most significantly influence the quality of reasoning. Such an investigation might lead to methodologies for measuring the DRI during the deliberation itself, potentially leading to an augmented form of DRI that doesn't require the participants to complete additional surveys. Real-time assessments could enhance the responsiveness of facilitation strategies and improve the overall effectiveness of deliberative forums.

A further area for exploration could be the integration of DRI measures as feedback or training parameters within machine learning and artificial intelligence models. By providing, for example, facilitation models, with data on how their outputs affect DRI outcomes, it would be possible to train them to promote behaviors that lead to increased deliberative quality. This integration could facilitate the development of AI-assisted tools that support participants in engaging more thoughtfully and constructively. Research in this domain would contribute to the advancement of digital deliberation platforms, making them more effective in fostering high-quality discourse and collective reasoning.

These recommendations underscore the potential for further innovation in the measurement and enhancement of deliberative processes. By leveraging technology and expanding data sources, future research can contribute to more efficient, inclusive, and insightful applications of the DRI. Such efforts would not only advance academic understanding but also have practical implications for the design and facilitation of deliberative democracy initiatives.

## 5. Conclusion

Our study set out to advance the measurement of deliberative processes by automating the generation of DRI statements. By integrating various forms of natural language processing and machine learning with human oversight, our approach mitigates the labor-intensive and subjective limitations inherent in traditional DRI survey generation significantly. The research has demonstrated that such an automated pipeline can efficiently process large-scale textual data from extensive news article corpora to extract, organize, and reformulate the core arguments embedded within public discourse. In doing so, we also contribute an innovative method for large-scale text analysis to the repertoire of computational social science.

The findings of this research demonstrate that AI and ML methods can effectively generate DRI survey statements and policy options by identifying and structuring public discourse at scale. The proposed method can significantly reduce the amount of human labor required for the generation of DRI surveys. By automating the extraction and categorization of key arguments, this method minimizes the need for extensive manual coding while maintaining high-quality output. The ability to process vast amounts of text from diverse sources improves the inclusivity and comprehensiveness of DRI statements. This approach further mitigates the risk of coder bias. The successful application to the 'Bevölkerungsrat 2025' citizens' assembly with minimal human involvement demonstrates the practical feasibility of the method. The validation results demonstrate a significant degree of overlap with human-generated DRI statements, underscoring the method's reliability. However, while effective, the methodology revealed certain limitations, namely the challenges in capturing minority viewpoints and the dependency on high-quality textual corpora. These issues highlight areas for future development.

Several avenues for future research emerge from this study. Expanding the dataset beyond news articles to include social media discussions could improve the method's representativeness and capture a wider range of perspectives. Further improvements in AI-driven text generation could enhance the specificity and diversity of DRI statements, reducing potential biases in framing and formulation. Exploring how AI and ML can be used to evaluate deliberation in real time, applying automated transcription and analysis of deliberative discussions, could provide additional insights into the dynamics of deliberation. Future research could investigate the integration of AI-generated deliberative feedback into digital deliberation platforms, helping facilitators improve discourse quality dynamically.

By demonstrating the feasibility of integrating AI- and ML-driven methods into public discourse analysis and the processing of large text corpora, this study makes a significant contribution to computational social science. The method offers an innovative fusion of computational text analysis with deliberative theory, providing new avenues for empirical research in deliberative democracy. The automation of DRI survey generation allows for broader and more frequent applications, enabling comparative and longitudinal studies on deliberative processes across different contexts. The method paves the way for AI-assisted facilitation tools in online deliberative platforms, enhancing the quality of discourse and participant engagement.

While some challenges remain, the integration of AI and ML into deliberative democracy research holds great promise for improving the inclusivity, efficiency, and impact of public deliberation. As computational methods continue to evolve, their application in deliberative practice will expand further, fostering a more nuanced and data-driven understanding of deliberative processes and enabling new avenues for citizen engagement.

Automatic Generation of DRI Statements                                    Maurice Flechtner

# Appendices

## A. Keywords & Prompts for keyword search

   I.   List of Keywords for news article selection:

| German | French | Italian |
| --- | --- | --- |
| Gesundheitskosten | Coûts de la santé | Costi della salute |
| Gesundheitssystem | Système de santé | Sistema sanitario |
| Gesundheitswesen | Système de santé | Settore sanitario |
| Gesundheitspolitik | Politique de santé | Politica sanitaria |
| Gesundheitsreform | Secteur de la santé | Riforma sanitaria |
| Gesundheitssektor | Réforme de la santé | Settore della salute |
| Gesundheitsversorgung | Secteur sanitaire | Accesso ai servizi sanitari |
| Spitalwesen | Accès aux soins de santé | Settore ospedaliero |
| Versicherungspflicht | Secteur hospitalier | Obbligo di assicurazione |
| Versicherungsmodell | Obligation d'assurance | Modello assicurativo |
| Krankenkassen | Modèle d'assurance | Casse malati |
| Krankenversicherung | Caisses maladie | Assicurazione malattia |
| Krankenhaus | Assurance maladie | Ospedale |
| Spital | Hôpital | Centro ospedaliero |
| Gesundheistprävention | Centre hospitalier | Prevenzione sanitaria |
| Gesundheitsförderung | Prévention sanitaire | Promozione della salute |
| Gesundheitsdienste | Promotion de la santé | Servizi sanitari |
|  | Services de santé |  |

## B. Anchor Statements & Prompts

I. Lists of anchor statements

General anchor statements

| German | French | Italian |
| --- | --- | --- |
| Die steigenden Gesundheitskosten in der Schweiz stellen eine komplexe Herausforderung dar, die durch verschiedene Faktoren beeinflusst wird. Das Krankenversicherungssystem spielt eine zentrale Rolle, da es die Finanzierung der Gesundheitsversorgung sichert, gleichzeitig aber auch durch die Struktur der obligatorischen Krankenpflegeversicherung (OKP) und den umfassenden Leistungskatalog Anreize für Mengenausweitungen schafft. Soziale Aspekte müssen beachtet werden, da die regressive Finanzierung über Kopfprämien einkommensschwächere Haushalte besonders belastet, obwohl Prämienverbilligungen eine gewisse Entlastung bieten. Im Bereich des Spitalwesens tragen die hohe Spitaldichte und die finanzielle Struktur der Spitalfinanzierung ebenfalls zur Kostensteigerung bei. Zudem ist die Gesundheitsversorgung oft fragmentiert, was zu ineffizienten Versorgungsabläufen führt und die Kosten weiter in die Höhe treibt. Schließlich wird das Potenzial der Gesundheitsprävention in der Schweiz noch immer nicht ausreichend genutzt, obwohl Prävention ein entscheidender Faktor sein könnte, um die langfristigen Gesundheitskosten zu senken. | Le système de santé suisse se caractérise par une grande complexité et une diversité d'acteurs, ce qui rend difficile la mise en œuvre de réformes non incrémentales, c'est-à-dire globales et profondes. L'analyse des politiques publiques montre que l'augmentation des coûts de la santé est influencée par une multitude de facteurs, dont l'augmentation du volume des traitements, le changement démographique et le développement technologique. Les propositions de réforme incluent une modification des modèles de rémunération, l'augmentation de la participation financière des patients et une possible réduction du catalogue des prestations de l'assurance-maladie obligatoire (LAMal). Cependant, ces approches entraînent d'importants conflits d'objectifs, notamment en ce qui concerne la justice sociale et l'égalité des chances dans l'accès aux services de santé. Le processus politique en Suisse, marqué par des structures fédérales et la démocratie directe, conduit souvent à des réformes progressives et incrémentales plutôt qu'à des changements systémiques globaux. | Il sistema sanitario svizzero si caratterizza per la grande complessità e la diversità degli attori, il che rende difficile l'attuazione di riforme non incrementali, ossia globali e profonde. L'analisi delle politiche pubbliche mostra che l'aumento dei costi sanitari è influenzato da una molteplicità di fattori, tra cui l'espansione del volume dei trattamenti, il cambiamento demografico e lo sviluppo tecnologico. Le proposte di riforma includono una modifica dei modelli di remunerazione, l'aumento della partecipazione finanziaria dei pazienti e una possibile riduzione del catalogo delle prestazioni dell'assicurazione obbligatoria delle cure medico-sanitarie (LAMal). Tuttavia, questi approcci comportano importanti conflitti di obiettivi, in particolare per quanto riguarda la giustizia sociale e l'uguaglianza di accesso ai servizi sanitari. Il processo politico in Svizzera, caratterizzato da strutture federali e democrazia diretta, porta spesso a riforme progressive e incrementali anziché a cambiamenti sistemici globali. |

Subcategory 1: Health system of Switzerland and non-incremental policy reforms

| German | French | Italian |
| --- | --- | --- |
| Das Schweizer Gesundheitssystem zeichnet sich durch eine hohe Komplexität und Vielfalt der Akteure aus, was es schwierig macht, nicht-inkrementelle, also umfassende und tiefgreifende Reformen durchzuführen. Die Politikfeldanalyse zeigt, dass die steigenden Gesundheitskosten von einer Vielzahl von Faktoren beeinflusst werden, darunter die Ausweitung der Behandlungsmenge, der demografische Wandel und die technologische Entwicklung. Reformvorschläge umfassen eine Änderung der Vergütungsmodelle, die Erhöhung der Eigenbeteiligung der Patienten und eine mögliche Reduktion des Leistungskatalogs der obligatorischen Krankenpflegeversicherung (OKP). Diese Ansätze bergen jedoch erhebliche Zielkonflikte, insbesondere hinsichtlich der sozialen Gerechtigkeit und der Chancengleichheit im Zugang zu Gesundheitsleistungen. Der politische Prozess in der Schweiz, geprägt durch föderale Strukturen und direkte Demokratie, führt oft dazu, dass Reformen schrittweise und inkrementell erfolgen, anstatt umfassende Systemveränderungen anzustreben. | Le système de santé suisse se caractérise par une grande complexité et une diversité d'acteurs, ce qui rend difficile la mise en œuvre de réformes non incrémentales, c'est-à-dire globales et profondes. L'analyse des politiques publiques montre que l'augmentation des coûts de la santé est influencée par une multitude de facteurs, dont l'augmentation du volume des traitements, le changement démographique et le développement technologique. Les propositions de réforme incluent une modification des modèles de rémunération, l'augmentation de la participation financière des patients et une possible réduction du catalogue des prestations de l'assurance-maladie obligatoire (LAMal). Cependant, ces approches entraînent d'importants conflits d'objectifs, notamment en ce qui concerne la justice sociale et l'égalité des chances dans l'accès aux services de santé. Le processus politique en Suisse, marqué par des structures fédérales et la démocratie directe, conduit souvent à des réformes progressives et incrémentales plutôt qu'à des changements systémiques globaux. | Il sistema sanitario svizzero si caratterizza per la grande complessità e la diversità degli attori, il che rende difficile l'attuazione di riforme non incrementali, ossia globali e profonde. L'analisi delle politiche pubbliche mostra che l'aumento dei costi sanitari è influenzato da una molteplicità di fattori, tra cui l'espansione del volume dei trattamenti, il cambiamento demografico e lo sviluppo tecnologico. Le proposte di riforma includono una modifica dei modelli di remunerazione, l'aumento della partecipazione finanziaria dei pazienti e una possibile riduzione del catalogo delle prestazioni dell'assicurazione obbligatoria delle cure medico-sanitarie (LAMal). Tuttavia, questi approcci comportano importanti conflitti di obiettivi, in particolare per quanto riguarda la giustizia sociale e l'uguaglianza di accesso ai servizi sanitari. Il processo politico in Svizzera, caratterizzato da strutture federali e democrazia diretta, porta spesso a riforme progressive e incrementali anziché a cambiamenti sistemici globali. |

Subcategory 2: Increasing utilization of healthcare services per person

| **German** | **French** | **Italian** |
|---|---|---|
| Die Politikfeldanalyse zur "steigenden Nutzung von Gesundheitsdienstleistungen pro Person" beleuchtet die zunehmende Inanspruchnahme von Gesundheitsleistungen in der Schweiz. Ein zentraler Faktor ist, dass die durchschnittlichen Bruttoleistungen pro versicherte Person von 2012 bis 2022 kontinuierlich gestiegen sind. Diese Entwicklung ist nicht allein durch die Alterung der Bevölkerung oder die Zunahme von Krankheiten zu erklären. Vielmehr wird ein Großteil des Anstiegs durch eine Mengenausweitung der erbrachten Leistungen verursacht, die sowohl angebots- als auch nachfrageseitig bedingt ist. Auf der Angebotsseite führen insbesondere Anreize zur Erbringung zusätzlicher Leistungen im Rahmen des Einzelleistungsvergütungssystems zu einer Kostensteigerung. Auf der Nachfrageseite trägt das Phänomen des "Moral Hazard" dazu bei, dass Versicherte tendenziell mehr Leistungen in Anspruch nehmen, als sie tatsächlich benötigen. Diese Entwicklungen führen zu einer signifikanten Erhöhung der Gesundheitskosten, was in der Politik zunehmend diskutiert und mit verschiedenen Reformvorschlägen, wie der Ablösung der Einzelleistungsvergütung oder der Einführung höherer Selbstbeteiligungen, adressiert wird. | L'analyse des politiques sur « l'augmentation de l'utilisation des services de santé par personne » examine l'accroissement continu du recours aux services de santé en Suisse. Un facteur clé est que les prestations brutes moyennes par personne assurée ont augmenté de manière continue entre 2012 et 2022. Cette évolution ne peut être expliquée uniquement par le vieillissement de la population ou par l'augmentation des maladies. Une grande partie de cette hausse est plutôt due à une expansion du volume des prestations fournies, causée à la fois par des facteurs liés à l'offre et à la demande. Du côté de l'offre, les incitations à fournir des prestations supplémentaires dans le cadre du système de rémunération à l'acte entraînent une augmentation des coûts. Du côté de la demande, le phénomène du « risque moral » pousse les assurés à consommer plus de prestations qu'ils n'en ont réellement besoin. Ces évolutions entraînent une augmentation significative des coûts de la santé, ce qui suscite de plus en plus de débats politiques et est abordé par diverses propositions de réforme, telles que le remplacement de la rémunération à l'acte ou l'introduction de contributions plus élevées des assurés. | L'analisi delle politiche sull'«aumento dell'utilizzo dei servizi sanitari per persona» esamina il crescente ricorso ai servizi sanitari in Svizzera. Un fattore chiave è che le prestazioni lorde medie per persona assicurata sono aumentate costantemente tra il 2012 e il 2022. Questa evoluzione non può essere spiegata solo dall'invecchiamento della popolazione o dall'aumento delle malattie. Gran parte di questo aumento è piuttosto dovuto a un'espansione del volume delle prestazioni fornite, causata sia da fattori legati all'offerta che alla domanda. Sul versante dell'offerta, gli incentivi a fornire prestazioni aggiuntive nel contesto del sistema di remunerazione a prestazione determinano un aumento dei costi. Sul versante della domanda, il fenomeno del «rischio morale» porta gli assicurati a usufruire di più prestazioni di quelle di cui hanno realmente bisogno. Queste evoluzioni comportano un aumento significativo dei costi sanitari, che è sempre più al centro del dibattito politico ed è affrontato da varie proposte di riforma, come la sostituzione della remunerazione a prestazione o l'introduzione di una maggiore partecipazione finanziaria degli assicurati. |

Subcategory 3: Hospital planning and financing

| German | French | Italian |
| --- | --- | --- |
| Die Politikfeldanalyse im Bereich "Hospital planning and financing" (Krankenhausplanung und -finanzierung) thematisiert die komplexen Herausforderungen und Zielkonflikte im Gesundheitswesen, insbesondere im Kontext steigender Gesundheitskosten. Ein zentrales Problem ist die hohe Spitaldichte in der Schweiz, die zu erheblichen Kostensteigerungen führt. Dies liegt unter anderem an der Vielzahl kleinerer Krankenhäuser, die ineffizient arbeiten und häufig nicht ausgelastet sind. Gleichzeitig spielt die Finanzierung eine kritische Rolle, da diese über verschiedene Regimes, wie die obligatorische Krankenpflegeversicherung (OKP), Zusatzversicherungen und staatliche Mittel, erfolgt. Die Analyse hebt die Bedeutung einer effektiven Krankenhausplanung hervor, um die Versorgung sicherzustellen und gleichzeitig die Kosten zu kontrollieren. Hierbei werden mögliche Reformansätze, wie die Zentralisierung der Spitalversorgung und die Einführung neuer Finanzierungsmodelle, diskutiert, wobei stets ein Balanceakt zwischen Kosteneffizienz, Versorgungsqualität und Zugänglichkeit der Gesundheitsdienste notwendig ist. | L'analyse des politiques publiques dans le domaine de la planification et du financement des hôpitaux met en lumière les défis complexes et les conflits d'objectifs dans le système de santé, en particulier dans le contexte de la hausse des coûts de santé. Un problème central est la densité élevée des hôpitaux en Suisse, ce qui entraîne une augmentation significative des coûts. Cela s'explique en partie par le grand nombre de petits hôpitaux qui fonctionnent de manière inefficace et sont souvent sous-utilisés. Parallèlement, le financement joue un rôle crucial, étant assuré par différents régimes, tels que l'assurance-maladie obligatoire (LAMal), les assurances complémentaires et les fonds publics. L'analyse souligne l'importance d'une planification hospitalière efficace pour garantir l'accès aux soins tout en maîtrisant les coûts. Des réformes potentielles, telles que la centralisation des services hospitaliers et l'introduction de nouveaux modèles de financement, sont discutées, en tenant toujours compte de la nécessité de trouver un équilibre entre l'efficacité des coûts, la qualité des soins et l'accessibilité des services de santé. | L'analisi delle politiche nel campo della pianificazione e del finanziamento degli ospedali mette in evidenza le sfide complesse e i conflitti di obiettivi nel sistema sanitario, in particolare nel contesto dell'aumento dei costi sanitari. Un problema centrale è l'elevata densità ospedaliera in Svizzera, che comporta un aumento significativo dei costi. Ciò è dovuto, in parte, al gran numero di piccoli ospedali che operano in modo inefficiente e che spesso non sono sfruttati appieno. Parallelamente, il finanziamento gioca un ruolo cruciale, essendo garantito da diversi regimi, come l'assicurazione obbligatoria delle cure medico-sanitarie (LAMal), le assicurazioni complementari e i fondi statali. L'analisi sottolinea l'importanza di una pianificazione ospedaliera efficace per garantire l'accesso alle cure mantenendo sotto controllo i costi. Vengono discusse potenziali riforme, come la centralizzazione dei servizi ospedalieri e l'introduzione di nuovi modelli di finanziamento, tenendo sempre presente la necessità di trovare un equilibrio tra l'efficienza dei costi, la qualità delle cure e l'accessibilità dei servizi sanitari. |

Subcategory 4: Design/Configuration/Organisation of compulsory health insurance

| German | French | Italian |
| --- | --- | --- |
| Die Politikfeldanalyse zur Gestaltung, Konfiguration und Organisation der obligatorischen Krankenversicherung (OKP) in der Schweiz beleuchtet die komplexe Struktur und die wesentlichen Herausforderungen dieses Systems. Das OKP-System, das auf einem Versicherungsobligatorium mit Aufnahmepflicht basiert, deckt einen umfassenden Leistungskatalog ab, der durch gesetzlich definierte Kriterien geregelt ist. Die Prämien sind nicht einkommensabhängig, sondern richten sich nach dem Wohnort und dem gewählten Versicherungsmodell, was zu einer regressiven Finanzierungsstruktur führt, bei der einkommensschwächere Haushalte proportional stärker belastet werden. Die Analyse zeigt auf, dass das System durch Fehlanreize wie die Mengenausweitung, insbesondere im ambulanten und stationären Bereich, unter Druck steht. Zudem wird der zunehmende Leistungskatalog als ein weiterer Kostentreiber identifiziert. Lösungsvorschläge umfassen unter anderem die Reform der Vergütungssysteme, eine stärkere Eigenbeteiligung der Versicherten und die Beschränkung des Leistungskatalogs, wobei jedoch Zielkonflikte und potenzielle negative Auswirkungen auf die Versicherten berücksichtigt werden müssen. | L'analyse des politiques concernant la conception, la configuration et l'organisation de l'assurance maladie obligatoire (AMO) en Suisse met en lumière la structure complexe et les principaux défis de ce système. Le système AMO, basé sur une obligation d'assurance avec obligation d'accepter tout assuré, couvre un large catalogue de prestations réglementé par des critères définis par la loi. Les primes ne sont pas ajustées en fonction des revenus, mais dépendent du lieu de résidence et du modèle d'assurance choisi, ce qui conduit à une structure de financement régressive, où les ménages à faible revenu sont proportionnellement plus lourdement chargés. L'analyse montre que le système est sous pression en raison de mauvaises incitations telles que l'augmentation du volume des prestations, en particulier dans les secteurs ambulatoire et hospitalier. De plus, l'élargissement du catalogue des prestations est identifié comme un autre facteur de coût. Parmi les propositions de solutions, on trouve la réforme des systèmes de rémunération, une participation plus forte des assurés aux coûts, et la limitation du catalogue des prestations, tout en tenant compte des conflits d'objectifs et des éventuels effets négatifs sur les assurés. | L'analisi delle politiche riguardanti la progettazione, la configurazione e l'organizzazione dell'assicurazione sanitaria obbligatoria (LAMal) in Svizzera mette in luce la complessa struttura e le principali sfide di questo sistema. Il sistema LAMal, basato su un obbligo di assicurazione con l'obbligo di accettare tutti gli assicurati, copre un ampio catalogo di prestazioni regolato da criteri definiti per legge. I premi non sono adattati in base al reddito, ma dipendono dal luogo di residenza e dal modello di assicurazione scelto, il che porta a una struttura di finanziamento regressiva, in cui le famiglie a basso reddito sono proporzionalmente più gravate. L'analisi mostra che il sistema è sotto pressione a causa di incentivi distorti, come l'aumento del volume delle prestazioni, in particolare nei settori ambulatoriale e ospedaliero. Inoltre, l'ampliamento del catalogo delle prestazioni è identificato come un ulteriore fattore di costo. Tra le proposte di soluzioni vi sono la riforma dei sistemi di remunerazione, una maggiore partecipazione dei cittadini ai costi e la limitazione del catalogo delle prestazioni, tenendo però conto dei conflitti di obiettivi e dei possibili effetti negativi sugli assicurati. |

Subcategory 5: Lack of incentives for health prevention

| German | French | Italian |
|---|---|---|
| Die fehlenden Anreize zur Gesundheitsprävention in der Schweiz resultieren aus einer komplexen Mischung aus strukturellen und politischen Faktoren. Obwohl Prävention langfristig die Gesundheitskosten senken könnte, gibt es auf individueller und institutioneller Ebene wenig Motivation, präventive Maßnahmen zu fördern. Die derzeitigen Anreizstrukturen im Gesundheitssystem begünstigen eher kurative Behandlungen als präventive Maßnahmen, da kurzfristige Einsparungen oft über langfristige Kostensenkungen gestellt werden. Zudem sind Präventionsprogramme häufig mit Zielkonflikten behaftet, beispielsweise zwischen der individuellen Freiheit und dem kollektiven Gesundheitsnutzen. Diese Zielkonflikte erschweren die politische Umsetzung von Präventionsmaßnahmen weiter und führen dazu, dass die Schweiz im europäischen Vergleich bei der Prävention zurückfällt. | Le manque d'incitations à la prévention en matière de santé en Suisse résulte d'un mélange complexe de facteurs structurels et politiques. Bien que la prévention puisse réduire les coûts de santé à long terme, il existe peu de motivation, tant au niveau individuel qu'institutionnel, pour promouvoir des mesures préventives. Les structures d'incitation actuelles du système de santé favorisent davantage les traitements curatifs que les mesures préventives, car les économies à court terme sont souvent privilégiées par rapport à la réduction des coûts à long terme. De plus, les programmes de prévention sont souvent confrontés à des conflits d'objectifs, par exemple entre la liberté individuelle et le bénéfice collectif pour la santé. Ces conflits rendent la mise en œuvre politique des mesures de prévention plus difficile, ce qui fait que la Suisse est à la traîne par rapport à d'autres pays européens en matière de prévention. | La mancanza di incentivi per la prevenzione sanitaria in Svizzera è il risultato di una combinazione complessa di fattori strutturali e politici. Sebbene la prevenzione possa ridurre i costi sanitari a lungo termine, esiste poca motivazione, sia a livello individuale che istituzionale, per promuovere misure preventive. Le attuali strutture di incentivo nel sistema sanitario favoriscono maggiormente i trattamenti curativi rispetto alle misure preventive, poiché i risparmi a breve termine sono spesso preferiti alla riduzione dei costi a lungo termine. Inoltre, i programmi di prevenzione sono spesso caratterizzati da conflitti di obiettivi, ad esempio tra la libertà individuale e il beneficio collettivo per la salute. Questi conflitti rendono più difficile l'attuazione politica delle misure di prevenzione, facendo sì che la Svizzera resti indietro rispetto ad altri paesi europei in materia di prevenzione. |

Subcategory 6: Coordination of health services

| **German** | **French** | **Italian** |
|---|---|---|
| Die mangelnde Koordination der Gesundheitsversorgung in der Schweiz stellt ein erhebliches Problem dar, insbesondere angesichts der steigenden Gesundheitskosten. Laut einer Politikfeldanalyse wird die koordinierte Versorgung als wesentliche Methode zur Verbesserung der Behandlungsqualität betrachtet, insbesondere für ältere Patienten mit chronischen Krankheiten, die von mehreren Fachpersonen betreut werden. Die Hausärzte spielen dabei eine zentrale Rolle als Gatekeeper, doch der zunehmende Mangel an Hausärzten verschärft die Problematik. Lösungsvorschläge umfassen unter anderem die Einführung von obligatorischen Gatekeeping-Modellen in der Grundversicherung sowie die Förderung der interprofessionellen Zusammenarbeit durch gezielte finanzielle Anreize. Ein Zielkonflikt besteht jedoch in der Einschränkung der Wahlfreiheit der Versicherten, was auf Widerstand stoßen könnte. | Le manque de coordination dans les services de santé en Suisse constitue un problème important, notamment en raison de l'augmentation des coûts de la santé. Selon une analyse des politiques, la coordination des soins est considérée comme une méthode essentielle pour améliorer la qualité des traitements, en particulier pour les patients âgés souffrant de maladies chroniques, qui sont suivis par plusieurs professionnels de santé. Les médecins généralistes jouent un rôle central en tant que gardiens du système, mais la pénurie croissante de généralistes aggrave le problème. Parmi les solutions proposées figurent l'introduction de modèles de garde obligatoires dans l'assurance de base et la promotion de la collaboration interprofessionnelle grâce à des incitations financières ciblées. Un conflit d'objectifs existe toutefois dans la limitation de la liberté de choix des assurés, ce qui pourrait rencontrer de la résistance. | La mancanza di coordinamento nei servizi sanitari in Svizzera rappresenta un problema significativo, soprattutto a causa dell'aumento dei costi sanitari. Secondo un'analisi delle politiche, il coordinamento dell'assistenza è considerato un metodo essenziale per migliorare la qualità delle cure, in particolare per i pazienti anziani con malattie croniche, che sono seguiti da diversi professionisti della salute. I medici di base svolgono un ruolo centrale come "gatekeeper" del sistema, ma la crescente carenza di medici di base aggrava il problema. Tra le soluzioni proposte figurano l'introduzione di modelli obbligatori di "gatekeeping" nell'assicurazione di base e la promozione della collaborazione interprofessionale attraverso incentivi finanziari mirati. Tuttavia, esiste un conflitto di obiettivi nella limitazione della libertà di scelta degli assicurati, che potrebbe incontrare resistenza. |

Subcategory 7: Financial burden of health costs and household (costs distribution)

| German | French | Italian |
| --- | --- | --- |
| Die Politikfeldanalyse zu den steigenden Gesundheitskosten beleuchtet die finanzielle Belastung, die durch Gesundheitskosten auf die Haushalte zukommt, und wie diese Kosten verteilt werden. In der Schweiz werden die Gesundheitskosten größtenteils durch die obligatorische Krankenpflegeversicherung (OKP), den Staat und die Haushalte selbst getragen. Haushalte sind sowohl über Prämien für die OKP und Zusatzversicherungen als auch durch Selbstzahlungen direkt an den Kosten beteiligt. Seit 2012 haben die Gesundheitskosten stark zugenommen, was zu einer höheren finanziellen Belastung der Haushalte führte, insbesondere für einkommensschwächere Familien. Der Anteil der Gesundheitsausgaben am Bruttohaushaltseinkommen ist regressiv, was bedeutet, dass einkommensschwächere Haushalte einen größeren Teil ihres Einkommens für Gesundheitsausgaben aufwenden müssen als einkommensstärkere. Diese Belastung wird nur teilweise durch staatliche Maßnahmen wie Prämienverbilligungen gemildert. Die steigenden Prämien und die zunehmende Notwendigkeit von Selbstzahlungen verschärfen die finanzielle Last der Haushalte und führen zu einer ungleichen Verteilung der Kosten im Gesundheitssystem. | L'analyse des politiques sur l'augmentation des coûts de la santé examine le fardeau financier que ces coûts imposent aux ménages et la manière dont ces dépenses sont réparties. En Suisse, les coûts de la santé sont principalement couverts par l'assurance-maladie obligatoire (LAMal), l'État et les ménages eux-mêmes. Les ménages contribuent directement aux dépenses de santé à travers les primes de la LAMal et des assurances complémentaires, ainsi que par les paiements directs. Depuis 2012, les coûts de la santé ont considérablement augmenté, ce qui a conduit à un fardeau financier plus important pour les ménages, en particulier pour les familles à faible revenu. La part des dépenses de santé dans le revenu brut des ménages est régressive, ce qui signifie que les ménages à faible revenu consacrent une plus grande part de leur revenu aux dépenses de santé que les ménages à revenu plus élevé. Ce fardeau n'est que partiellement atténué par des mesures étatiques telles que les réductions de primes. L'augmentation des primes et la nécessité croissante de paiements directs aggravent la charge financière des ménages et entraînent une répartition inégale des coûts dans le système de santé. | L'analisi delle politiche sui crescenti costi sanitari esamina il peso finanziario che questi costi impongono alle famiglie e come tali spese vengono distribuite. In Svizzera, i costi sanitari sono principalmente coperti dall'assicurazione sanitaria obbligatoria (LAMal), dallo Stato e dalle famiglie stesse. Le famiglie contribuiscono direttamente alle spese sanitarie attraverso i premi della LAMal e delle assicurazioni complementari, nonché tramite pagamenti diretti. Dal 2012, i costi sanitari sono aumentati considerevolmente, portando a un maggiore onere finanziario per le famiglie, in particolare per quelle a basso reddito. La quota delle spese sanitarie rispetto al reddito lordo delle famiglie è regressiva, il che significa che le famiglie a basso reddito dedicano una parte maggiore del loro reddito alle spese sanitarie rispetto alle famiglie con redditi più alti. Questo onere è solo parzialmente attenuato da misure statali come le riduzioni dei premi. L'aumento dei premi e la crescente necessità di pagamenti diretti aggravano il carico finanziario delle famiglie e portano a una distribuzione diseguale dei costi nel sistema sanitario. |

II.   Prompts provided to the language model for anchor statement generation:

General anchor statement

Original prompt in German:

> "Du bist ein erfahrener Journalist, schreibe einen Absatz über die steigenden Gesundheitskosten in der Schweiz und welche Aspekte man dabei beachten muss. Nutze das angehangene Dokument als Grundlage. Erwähne folgende Themen: Krankenversicherungssystem, soziale Aspekte, Spitalwesen, Gesundheitsversorgung, Gesundheitsprävention."

Translation to English:

> "You are an experienced journalist. Write a paragraph about the rising healthcare costs in Switzerland and the aspects that need to be considered. Use the attached document as a basis. Mention the following topics: health insurance system, social aspects, hospital sector, healthcare provision, and health prevention."

We appended the Politikfeldanalyse by Interface to provide context. The output was then translated to French and Italian.

Anchor statements for subcategories

Original prompt in German:

> "Du bist ein erfahrener Journalist, schreibe einen zusammenfassenden Absatz über '[insert subcategory here]' auf Basis der Politikfeldanalyse."

Translation to English:

> "You are an experienced journalist. Write a summary paragraph about '[insert subcategory here]' based on the policy analysis."

We appended the Politikfeldanalyse by Interface to provide context. The output was then translated to French and Italian.

List of subcategories inserted:

| 1. | Health system of Switzerland and non-incremental policy reforms |
|----|------------------------------------------------------------------|
| 2. | Increasing utilisation of healthcare services per person |
| 3. | Hospital planning and financing |
| 4. | Design/Configuration/Organisation of Compulsory health insurance |
| 5. | Lack of incentives for Health Prevention |
| 6. | Coordination of health services |
| 7. | Financial burden of health costs and household (costs distribution) |

**C. Validation**

  I. List of keywords for news article selection for validation

| **German** | **French** | **Italian** |
| --- | --- | --- |
| Lebensmittelsystem | Système alimentaire | Sistema alimentare |
| Nachhaltigkeit | Durabilité | Sostenibilità |
| Landwirtschaft | Agriculture | Agricoltura |
| Klimawandel | Climat | Clima |
| Agrarpolitik | Politique agricole | Politica agricola |
| Biodiversität | Biodiversité | Biodiversità |
| Lebensmittelsicherheit | Sécurité alimentaire | Sicurezza alimentare |
| Regionalität | Localité | Località |
| Agroökologie | Agroécologie | Agroecologia |
| Pestizidfreiheit | Pesticides | Pesticidi |
| Ernährung | Alimentation | Alimentazione |
| Treibhausgas | Gaz à effet de serre | Gas serra |
| Biolandbau | Agriculture biologique | Agricoltura biologica |
| Importabhängigkeit | Importation | Importazioni |
| Ernährungspolitik | Politique alimentaire | Politica alimentare |

II. List of anchor statements for validation

General anchor statements

| **German** | **French** | **Italian** |
|---|---|---|
| Das Schweizer Ernährungssystem steht an einem Wendepunkt und benötigt eine durchdachte Transformation, um ökologischen, wirtschaftlichen und sozialen Herausforderungen zu begegnen. Der Wandel ist unausweichlich, da die aktuellen Systeme mit Problemen wie Umweltverschmutzung, Biodiversitätsverlust und Ineffizienzen kämpfen, die den Nachhaltigkeitszielen entgegenstehen. Die Verantwortung für Maßnahmen liegt bei verschiedenen Akteuren: Regierungen müssen strategische Eingriffe vornehmen, etwa durch die Integration von Umweltkriterien in landwirtschaftliche Subventionen, während Unternehmen und Verbraucher gemeinsam die Pflicht tragen, Abfall zu reduzieren und auf nachhaltige Praktiken umzustellen. Wirtschaftlich gesehen ist die Schweizer Landwirtschaft sowohl global wettbewerbsfähig als auch stark geschützt, was ausgewogene Politiken erfordert, die Innovationen fördern, ohne Isolationismus zu begünstigen. Darüber hinaus erfordern Umweltverschmutzung und der Verlust der Biodiversität dringend Maßnahmen, da intensive landwirtschaftliche Praktiken die natürlichen Lebensräume erheblich geschädigt haben. Eine nachhaltige Landwirtschaft, die auf ressourcenschonende und umweltfreundliche Methoden setzt, bietet einen gangbaren Weg, benötigt jedoch systematische Unterstützung und marktbasierte Anreize. Aus | Le système alimentaire suisse se trouve à un tournant et nécessite une transformation réfléchie pour relever les défis écologiques, économiques et sociaux. Le changement est inévitable, car les systèmes actuels peinent à résoudre des problèmes tels que la pollution, la perte de biodiversité et les inefficacités qui vont à l'encontre des objectifs de durabilité. La responsabilité de l'action incombe à de multiples acteurs : les gouvernements doivent intervenir de manière stratégique, par exemple en intégrant des critères environnementaux dans les subventions agricoles, tandis que les entreprises et les consommateurs partagent le devoir de réduire les déchets et d'adopter des pratiques durables. Économiquement, l'agriculture suisse est à la fois compétitive à l'échelle mondiale et fortement protégée, ce qui nécessite des politiques équilibrées qui soutiennent l'innovation sans encourager l'isolationnisme. Par ailleurs, la pollution et la perte de biodiversité exigent des réponses urgentes, car les pratiques agricoles intensives ont considérablement dégradé les habitats naturels. Une agriculture durable, qui privilégie des méthodes éco-responsables et à faible impact, offre une voie viable, mais nécessite un soutien systématique et des incitations basées sur le marché. D'un point de vue global, la Suisse doit également | Il sistema alimentare svizzero si trova a un punto di svolta e necessita di una trasformazione ponderata per affrontare le sfide ecologiche, economiche e sociali. Il cambiamento è inevitabile, poiché i sistemi attuali faticano a risolvere problemi come l'inquinamento, la perdita di biodiversità e le inefficienze che ostacolano gli obiettivi di sostenibilità. La responsabilità dell'azione ricade su molteplici attori: i governi devono intervenire strategicamente, ad esempio integrando criteri ambientali nei sussidi agricoli, mentre imprese e consumatori condividono il dovere di ridurre gli sprechi e adottare pratiche sostenibili. Dal punto di vista economico, l'agricoltura svizzera è sia competitiva a livello globale che fortemente protetta, richiedendo politiche equilibrate che promuovano l'innovazione senza favorire l'isolazionismo. Inoltre, l'inquinamento e la perdita di biodiversità richiedono risposte urgenti, poiché le pratiche agricole intensive hanno gravemente degradato gli habitat naturali. L'agricoltura sostenibile, che privilegia metodi ecologici e a basso impatto, rappresenta una strada percorribile, ma necessita di un supporto |
| globaler Perspektive muss die Schweiz sich internationalen Standards anpassen, Nachhaltigkeit in Handelsabkommen fördern und ihren Einfluss nutzen, um weltweit umweltbewusste Praktiken voranzutreiben. Dieser Wandel kann nur durch koordinierte Maßnahmen und entschlossene Antworten von Politik, Märkten und Gesellschaft gelingen, damit Ernährungssysteme sowohl die Menschen als auch den Planeten nachhaltig versorgen können. | s'aligner sur les normes internationales, promouvoir la durabilité dans les accords commerciaux et utiliser son influence pour encourager des pratiques écologiques à travers le monde. Cette transformation ne pourra réussir que grâce à une action coordonnée et des réponses déterminées des décideurs politiques, des marchés et de la société, garantissant ainsi que les systèmes alimentaires puissent soutenir à la fois les populations et la planète de manière durable. | sistematico e di incentivi di mercato. Da una prospettiva globale, la Svizzera deve anche allinearsi agli standard internazionali, promuovere la sostenibilità negli accordi commerciali e utilizzare la sua influenza per incoraggiare pratiche ecologiche in tutto il mondo. Questa trasformazione potrà avere successo solo attraverso un'azione coordinata e risposte decise da parte dei decisori politici, dei mercati e della società, garantendo che i sistemi alimentari possano sostenere in modo duraturo sia le persone che il pianeta. |

Subcategory 1: Reality of Change

| German | French | Italian |
| --- | --- | --- |
| Die Transformation des Schweizer Ernährungssystems spiegelt ein komplexes Zusammenspiel zwischen der Anerkennung von Krisen und der Überwindung von Veränderungsbarrieren wider. Das System steht vor einem „unaufgelösten Steuerungsproblem", bei dem produktivistische Politiken historisch daran gescheitert sind, den weltweiten Hunger und ökologische Herausforderungen zu bewältigen. Dies hat Unterernährung, Fehlernährung und Überernährung zusätzlich verschärft. Basisbewegungen und globale Initiativen machen auf diese Krisen aufmerksam, doch bremsen die Macht großer Konzerne, neoliberale Einflüsse und die Priorisierung von Expertenwissen gegenüber dem Wissen der Allgemeinheit den Fortschritt. Jüngste Bemühungen, wie der Schweizer Bürger:innenrat für Ernährungspolitik, setzen auf inklusive, demokratische Beratungen, um Nachhaltigkeit mit den Realitäten von Produktion und Konsum in Einklang zu bringen. Dieser Wandel unterstreicht die Dringlichkeit, gesellschaftliche Erwartungen mit ökologischen und wirtschaftlichen Realitäten zu vereinen, um eine widerstandsfähige und gerechte Ernährungszukunft zu erreichen | La transformation du système alimentaire suisse reflète une interaction complexe entre la reconnaissance des crises et le dépassement des obstacles au changement. Le système est confronté à un « problème de gouvernance non résolu », où les politiques productivistes ont historiquement échoué à résoudre la faim mondiale et les défis environnementaux, aggravant ainsi la sous-alimentation, la malnutrition et la suralimentation. Les mouvements citoyens et les initiatives globales mettent en lumière ces crises, mais la puissance des grandes entreprises, les influences néolibérales et la priorisation du savoir expert au détriment des connaissances citoyennes freinent les progrès. Les efforts récents, comme l'Assemblée citoyenne suisse pour la politique alimentaire, misent sur des délibérations inclusives et démocratiques pour aligner la durabilité sur les réalités de la production et de la consommation. Ce changement souligne l'urgence de concilier les attentes sociétales avec les réalités écologiques et économiques afin de parvenir à un avenir alimentaire résilient et équitable | La trasformazione del sistema alimentare svizzero riflette un complesso intreccio tra il riconoscimento delle crisi e il superamento degli ostacoli al cambiamento. Il sistema affronta un "problema di governance irrisolto", in cui le politiche produttivistiche hanno storicamente fallito nel risolvere la fame globale e le sfide ambientali, aggravando la sottoalimentazione, la malnutrizione e la sovralimentazione. Movimenti di base e iniziative globali evidenziano queste crisi, ma il potere delle grandi aziende, le influenze neoliberiste e la priorità data al sapere degli esperti rispetto a quello dei cittadini rallentano i progressi. Sforzi recenti, come l'Assemblea dei cittadini svizzeri per la politica alimentare, puntano su consultazioni democratiche e inclusive per allineare la sostenibilità con le realtà della produzione e del consumo. Questo cambiamento evidenzia l'urgenza di conciliare le aspettative sociali con le realtà ecologiche ed economiche per raggiungere un futuro alimentare resiliente ed equo |

Subcategory 2: Responsibility of Action

| German | French | Italian |
| --- | --- | --- |
| Das Schweizer Lebensmittelsystem steht vor einem dringenden Bedarf an Transformation, der sowohl durch systemische Krisen als auch durch gesellschaftliche Erwartungen nach mehr Nachhaltigkeit, Fairness und ökologischer Verantwortung getrieben wird. Zu den aktuellen Herausforderungen gehören hohe Agrarsubventionen, eine begrenzte Selbstversorgung und ökologische Belastungen durch intensive Landwirtschaftspraktiken. Basisbewegungen und Bürgerinitiativen verdeutlichen die wachsende öffentliche Forderung nach Reformen, wie der Reduzierung des Pestizideinsatzes und der Förderung von biologischer Landwirtschaft. Dennoch bestehen systemische Hürden, darunter die Macht von Großkonzernen, unausgewogene politische Maßnahmen und fragmentierte Marktstrukturen. Der Bürger:innenrat für Ernährungspolitik in der Schweiz ist ein Beispiel für Bemühungen, einen inklusiven Dialog zu fördern und innovative, demokratische Lösungen zu entwickeln. Dieser Ansatz steht im Einklang mit der Schweizer Strategie für nachhaltige Entwicklung, die auf Zusammenarbeit zwischen Bürger:innen, politischen Entscheidungsträgern und Expert:innen setzt, um wirtschaftliche Tragfähigkeit, ökologische Gesundheit und gesellschaftliches Wohlergehen in Einklang zu bringen. | Le système alimentaire suisse fait face à un besoin urgent de transformation, motivé à la fois par des crises systémiques et par les attentes sociétales en matière de durabilité, d'équité et de responsabilité environnementale. Les défis actuels incluent des subventions agricoles élevées, une autosuffisance limitée et des pressions écologiques liées à des pratiques agricoles intensives. Les mouvements de base et les initiatives citoyennes reflètent la demande croissante de réformes, telles que la réduction de l'utilisation des pesticides et la promotion de l'agriculture biologique. Cependant, des obstacles systémiques persistent, notamment le pouvoir des grandes entreprises, des politiques désalignées et des structures de marché fragmentées. L'Assemblée citoyenne pour la politique alimentaire en Suisse illustre les efforts visant à encourager un dialogue inclusif et à développer des solutions démocratiques innovantes. Cette approche s'aligne sur la stratégie suisse de développement durable, mettant l'accent sur la collaboration entre citoyens, décideurs politiques et experts pour équilibrer viabilité économique, santé écologique et bien-être sociétal. | Il sistema alimentare svizzero affronta un bisogno urgente di trasformazione, guidato sia da crisi sistemiche che dalle aspettative della società in termini di sostenibilità, equità e responsabilità ambientale. Le sfide attuali includono alti sussidi agricoli, un'autosufficienza limitata e pressioni ecologiche derivanti da pratiche agricole intensive. I movimenti di base e le iniziative dei cittadini evidenziano una crescente richiesta di riforme, come la riduzione dell'uso di pesticidi e la promozione dell'agricoltura biologica. Tuttavia, persistono ostacoli sistemici, tra cui il potere delle grandi aziende, politiche disallineate e strutture di mercato frammentate. L'Assemblea dei Cittadini per la Politica Alimentare in Svizzera rappresenta un esempio di sforzi volti a promuovere un dialogo inclusivo e sviluppare soluzioni democratiche innovative. Questo approccio è in linea con la strategia svizzera per lo sviluppo sostenibile, che pone l'accento sulla collaborazione tra cittadini, decisori politici ed esperti per bilanciare sostenibilità economica, salute ecologica e benessere sociale. |

Subcategory 3: Economy and Markets

| German | French | Italian |
| --- | --- | --- |
| Die Zukunft des Schweizer Ernährungssystems im Bereich „Wirtschaft und Märkte" steht vor erheblichen Herausforderungen, birgt jedoch auch einzigartige Chancen. Der Agrarsektor ist im internationalen Vergleich stark geschützt, kämpft jedoch mit geringer Rentabilität, insbesondere in Bergregionen, wo die Einkommen bis zu 40 % unter vergleichbaren Löhnen liegen können. Das Schweizer Ernährungssystem steht in einem Spannungsfeld zwischen der Aufrechterhaltung der internationalen Wettbewerbsfähigkeit und der Erreichung von Nachhaltigkeitszielen. Innovationen wie gezielte Direktzahlungen und regulatorische Rahmenbedingungen sind notwendig, um ökologische Auswirkungen mit wirtschaftlicher Tragfähigkeit in Einklang zu bringen. Zudem bedarf es einer gerechteren Gewinnverteilung entlang der gesamten Wertschöpfungskette, um die Produzenten gegenüber der Marktmacht großer Verarbeiter und Einzelhändler zu stärken. Während die Schweizer Agrarpolitik Multifunktionalität und nachhaltige Entwicklung betont, wächst der Bedarf an Reformen, die die Erwartungen der Konsumenten, die Realitäten des internationalen Handels und den Schutz der Biodiversität miteinander in Einklang bringen. Diese Faktoren erfordern einen integrierten Ansatz in der Politikgestaltung, um eine widerstandsfähige und gerechte Ernährungswirtschaft zu fördern. | L'avenir du système alimentaire suisse dans le domaine de l'« économie et des marchés » fait face à d'importants défis tout en offrant des opportunités uniques. Le secteur agricole est fortement protégé par rapport aux normes internationales, mais il souffre d'une faible rentabilité, notamment dans les régions montagneuses où les revenus peuvent être inférieurs de 40 % aux salaires comparables. Le système alimentaire suisse se trouve dans une tension entre le maintien de la compétitivité internationale et l'atteinte des objectifs de durabilité. Des innovations telles que des paiements directs ciblés et des cadres réglementaires sont nécessaires pour équilibrer les impacts écologiques avec la viabilité économique. En outre, une répartition plus équitable des bénéfices tout au long de la chaîne de valeur est indispensable pour renforcer le pouvoir des producteurs face à la concentration du marché détenue par les grands transformateurs et détaillants. Bien que la politique agricole suisse mette l'accent sur la multifonctionnalité et le développement durable, la nécessité de réformes s'accroît pour aligner les attentes des consommateurs, les réalités du commerce international et la préservation de la biodiversité. Ces facteurs exigent une approche intégrée de la politique afin de promouvoir une économie alimentaire résiliente et équitable. | Il futuro del sistema alimentare svizzero nell'ambito di "Economia e Mercati" deve affrontare sfide significative ma offre anche opportunità uniche. Il settore agricolo è altamente protetto rispetto agli standard internazionali, ma lotta con una bassa redditività, in particolare nelle regioni montane, dove i redditi possono essere inferiori del 40% rispetto ai salari comparabili. Il sistema alimentare svizzero si trova in una tensione tra il mantenimento della competitività internazionale e il raggiungimento degli obiettivi di sostenibilità. Sono necessarie innovazioni, come pagamenti diretti mirati e quadri normativi, per bilanciare l'impatto ecologico con la sostenibilità economica. Inoltre, è essenziale una distribuzione più equa dei profitti lungo tutta la filiera, rafforzando il potere dei produttori contro la concentrazione del mercato detenuta dai grandi trasformatori e rivenditori. Sebbene la politica agricola svizzera enfatizzi la multifunzionalità e lo sviluppo sostenibile, cresce la necessità di riforme che allineino le aspettative dei consumatori, le realtà del commercio internazionale e la salvaguardia della biodiversità. |

| | | Questi fattori richiedono un approccio integrato alle politiche per promuovere un'economia alimentare resiliente ed equa. |
|---|---|---|

Subcategory 4: Pollution and Biodiversity

| **German** | **French** | **Italian** |
|---|---|---|
| Das Schweizer Ernährungssystem steht vor einer entscheidenden Transformation, wobei Verschmutzung und Biodiversität zentrale Herausforderungen darstellen. Die Landwirtschaft des Landes hat erhebliche Auswirkungen auf die Ökosysteme, insbesondere durch übermäßige Stickstoff- und Phosphornutzung, die den Verlust der Biodiversität und die Verschmutzung von Gewässern vorantreiben. Intensivierte Anbaumethoden, Monokulturen und eine hohe Viehdichte verschärfen diese Belastungen und gefährden das fragile Gleichgewicht der Agrarlandschaften. Um diese Probleme anzugehen, setzt die Schweiz auf Strategien wie die Förderung nachhaltiger Produktionspraktiken, die Schaffung von Anreizen für ökologische Maßnahmen und die Reduktion des Pestizideinsatzes. Die Politik strebt an, wirtschaftliche, ökologische und soziale Prioritäten in Einklang zu bringen, Innovationen zu fördern und gleichzeitig langfristige Ernährungssicherheit und ökologische Resilienz sicherzustellen. | Le système alimentaire suisse est à un tournant crucial, avec la pollution et la biodiversité au cœur des enjeux. L'agriculture du pays a des impacts significatifs sur les écosystèmes, notamment en raison de l'utilisation excessive d'azote et de phosphore, qui accélère la perte de biodiversité et la pollution des eaux. Les méthodes agricoles intensives, les monocultures et la forte densité de bétail aggravent ces pressions, menaçant l'équilibre fragile des paysages agricoles. Pour relever ces défis, la Suisse mise sur des stratégies telles que la promotion de pratiques de production durable, la création d'incitations pour des mesures écologiques et la réduction de l'utilisation des pesticides. Les décideurs politiques visent à équilibrer les priorités économiques, environnementales et sociales, à encourager l'innovation tout en garantissant la sécurité alimentaire et la résilience écologique à long terme. | Il sistema alimentare svizzero si trova a un punto di svolta cruciale, con l'inquinamento e la biodiversità al centro delle sfide. L'agricoltura del paese ha un impatto significativo sugli ecosistemi, in particolare a causa dell'uso eccessivo di azoto e fosforo, che accelera la perdita di biodiversità e la contaminazione delle acque. Le pratiche agricole intensive, le monocolture e l'elevata densità di bestiame aggravano queste pressioni, mettendo a rischio il delicato equilibrio dei paesaggi agricoli. Per affrontare queste problematiche, la Svizzera punta su strategie come la promozione di pratiche di produzione sostenibile, la creazione di incentivi per misure ecologiche e la riduzione dell'uso di pesticidi. I responsabili politici mirano a bilanciare le priorità economiche, ambientali e sociali, promuovendo l'innovazione e garantendo al contempo la sicurezza alimentare e la resilienza ecologica a lungo termine |

Subcategory 5: Sustainable Agriculture

| **German** | **French** | **Italian** |
|---|---|---|
| Die Zukunft der nachhaltigen Landwirtschaft in der Schweiz ist geprägt von einem komplexen Zusammenspiel aus ökologischen, wirtschaftlichen und sozialen Faktoren. Das Land steht vor der doppelten Herausforderung, die Produktivität zu steigern und gleichzeitig ökologische Grenzen einzuhalten, wobei der Agrarsektor weiterhin stark von Subventionen und Schutzmaßnahmen abhängig ist. Innovative Ansätze wie Bürger:innenräte und gezielte Direktzahlungen für umweltfreundliche Praktiken wurden vorgeschlagen, um systemische Barrieren zu überwinden. Basisbewegungen und globale Trends betonen den reduzierten Einsatz von Pestiziden, die Förderung der Biodiversität und eine verstärkte Orientierung hin zu pflanzlichen Ernährungsweisen. Angesichts einer breiten öffentlichen Unterstützung für Nachhaltigkeit, aber einer langsamen politischen Umsetzung, setzt die Schweiz auf eine Strategie, die protektionistische Tendenzen mit globaler Wettbewerbsfähigkeit ausbalanciert und offene Dialoge fördert, um ihre Ziele für nachhaltige Entwicklung zu erreichen. | L'avenir de l'agriculture durable en Suisse est marqué par une interaction complexe entre des facteurs écologiques, économiques et sociaux. Le pays doit relever le double défi d'augmenter la productivité tout en respectant les limites écologiques, alors que le secteur agricole reste fortement dépendant des subventions et des politiques de protection. Des approches innovantes, telles que les assemblées citoyennes et les paiements directs ciblés pour des pratiques respectueuses de l'environnement, ont été proposées pour surmonter les obstacles systémiques. Les mouvements de base et les tendances mondiales mettent en avant la réduction de l'utilisation des pesticides, la promotion de la biodiversité et une transition vers une alimentation davantage axée sur les végétaux. Face à un large soutien public pour la durabilité mais une mise en œuvre politique lente, la Suisse mise sur une stratégie visant à équilibrer les tendances protectionnistes avec la compétitivité mondiale et à favoriser les dialogues ouverts pour atteindre ses objectifs de développement durable. | Il futuro dell'agricoltura sostenibile in Svizzera è caratterizzato da un'interazione complessa tra fattori ecologici, economici e sociali. Il paese si trova ad affrontare la doppia sfida di aumentare la produttività rispettando i limiti ecologici, mentre il settore agricolo continua a dipendere fortemente da sovvenzioni e politiche protezionistiche. Sono stati proposti approcci innovativi, come le assemblee cittadine e i pagamenti diretti mirati per pratiche rispettose dell'ambiente, per superare le barriere sistemiche. I movimenti di base e le tendenze globali sottolineano la necessità di ridurre l'uso di pesticidi, promuovere la biodiversità e orientarsi verso un'alimentazione più vegetale. Di fronte a un ampio sostegno pubblico per la sostenibilità ma a un'implementazione politica lenta, la Svizzera punta su una strategia che bilanci le tendenze protezionistiche con la competitività globale e favorisca dialoghi aperti per raggiungere i suoi obiettivi di sviluppo sostenibile. |

Subcategory 6: Global Perspective

| German | French | Italian |
| --- | --- | --- |
| Die Zukunft des Schweizer Ernährungssystems steht vor erheblichen Herausforderungen und Chancen aus globaler Perspektive. Der landwirtschaftliche Rahmen des Landes ist stark geschützt, jedoch wirtschaftlich belastet, da die Einkommen der Landwirte deutlich unter vergleichbaren Löhnen liegen. Globale Einflüsse wie der Klimawandel, der Verlust der Biodiversität und die Nachhaltigkeit von Ernährungssystemen überschneiden sich mit lokalen Herausforderungen wie der Importabhängigkeit und der begrenzten Selbstversorgung. Die Schweizer Strategie legt Wert auf Multifunktionalität, Innovation und internationale Zusammenarbeit und positioniert ihre Ernährungspolitik als Modell für Nachhaltigkeit. Allerdings erschweren tief verwurzelte Barrieren, wie die Macht großer Unternehmen und Kommunikationskrisen, transformative Fortschritte. Basisbewegungen und Bürgerversammlungen spielen eine entscheidende Rolle bei der Förderung inklusiver, systemischer Veränderungen. Der Ausgleich zwischen Protektionismus und Offenheit sowie zwischen ökologischen Zielen und wirtschaftlicher Rentabilität wird entscheidend sein, um ein widerstandsfähiges und global bewusstes Ernährungssystem zu erreichen. | L'avenir du système alimentaire suisse fait face à des défis importants et à des opportunités dans une perspective globale. Le cadre agricole du pays est fortement protégé, mais économiquement sous pression, avec des revenus agricoles nettement inférieurs aux salaires comparables. Les influences mondiales, telles que le changement climatique, la perte de biodiversité et la durabilité des systèmes alimentaires, se croisent avec des contraintes locales, notamment la dépendance aux importations et une autosuffisance limitée. La stratégie suisse met l'accent sur la multifonctionnalité, l'innovation et la coopération internationale, positionnant ainsi sa politique alimentaire comme un modèle de durabilité. Cependant, des obstacles profondément ancrés, tels que le pouvoir des grandes entreprises et les crises de communication, limitent les avancées transformatrices. Les mouvements de base et les assemblées citoyennes jouent un rôle crucial dans la promotion de changements systémiques inclusifs. Trouver un équilibre entre protectionnisme et ouverture, ainsi qu'entre objectifs écologiques et viabilité économique, sera essentiel pour construire un système alimentaire résilient et conscient des enjeux mondiaux. | Il futuro del sistema alimentare svizzero affronta sfide significative e opportunità da una prospettiva globale. Il quadro agricolo del paese è altamente protetto ma economicamente sotto pressione, con redditi agricoli significativamente inferiori ai salari comparabili. Influenze globali, come il cambiamento climatico, la perdita di biodiversità e la sostenibilità dei sistemi alimentari, si intrecciano con sfide locali, tra cui la dipendenza dalle importazioni e una limitata autosufficienza. La strategia svizzera pone l'accento sulla multifunzionalità, l'innovazione e la cooperazione internazionale, posizionando la sua politica alimentare come un modello di sostenibilità. Tuttavia, ostacoli radicati, come il potere delle grandi aziende e le crisi di comunicazione, limitano i progressi trasformativi. I movimenti di base e le assemblee cittadine svolgono un ruolo cruciale nel promuovere cambiamenti sistemici inclusivi. Bilanciare protezionismo e apertura, così come obiettivi ecologici e sostenibilità economica, sarà fondamentale per realizzare un sistema alimentare resiliente e consapevole a livello globale. |

Subcategory 7: Action and Responses

| German | French | Italian |
| --- | --- | --- |
| Das Schweizer Ernährungssystem steht an einem Wendepunkt, geprägt von komplexen Herausforderungen und dynamischen Reaktionen. Angesichts eines wachsenden öffentlichen Bewusstseins für ökologische Grenzen und nachhaltige Praktiken sind Initiativen wie der Schweizer Bürger:innenrat für Ernährungspolitik entstanden, um inklusive Beratungen und politische Reformen voranzutreiben. Diese Versammlungen unterstreichen die Notwendigkeit, Basisbewegungen, Expertenwissen und staatliche Interventionen zu verbinden, um strukturelle Hürden wie die Dominanz von Konzernen und die Abhängigkeit von produktivistischen Paradigmen zu überwinden. Durch den Einsatz von bürgergeführtem Dialog und nachhaltigen Strategien strebt die Schweiz ein Ernährungssystem an, das Ernährungssicherheit, ökologische Nachhaltigkeit und soziale Gerechtigkeit in Einklang bringt und so ein zukunftsfähiges System schafft. | Le système alimentaire suisse se trouve à un carrefour décisif, marqué par des défis complexes et des réponses dynamiques. Face à une prise de conscience publique croissante des limites écologiques et des pratiques durables, des initiatives comme l'Assemblée citoyenne suisse pour la politique alimentaire ont vu le jour pour favoriser des délibérations inclusives et des réformes politiques. Ces assemblées soulignent la nécessité de relier les mouvements citoyens, les savoirs experts et les interventions étatiques afin de surmonter les obstacles structurels tels que la domination des entreprises et la dépendance aux paradigmes productivistes. En adoptant un dialogue citoyen et des stratégies durables, la Suisse vise à construire un système alimentaire qui équilibre sécurité alimentaire, durabilité environnementale et équité sociale, ouvrant ainsi la voie à un avenir résilient. | Il sistema alimentare svizzero si trova a un punto di svolta cruciale, caratterizzato da sfide complesse e risposte dinamiche. Di fronte a una crescente consapevolezza pubblica dei limiti ecologici e delle pratiche sostenibili, sono emerse iniziative come l'Assemblea dei cittadini svizzeri per la politica alimentare, volte a promuovere deliberazioni inclusive e riforme politiche. Queste assemblee evidenziano la necessità di collegare i movimenti di base, le conoscenze esperte e le politiche governative per superare ostacoli strutturali come il dominio delle grandi aziende e la dipendenza dai paradigmi produttivistici. Adottando il dialogo guidato dai cittadini e strategie sostenibili, la Svizzera punta a costruire un sistema alimentare che equilibri sicurezza alimentare, sostenibilità ambientale ed equità sociale, tracciando così un percorso verso un futuro resiliente. |

III.   Prompts provided to the language model for anchor statement generation:

   General anchor statement

   Original prompt:

   > "You are an experienced journalist. Write a paragraph about the future of Switzerland's food system and all aspects that need to be considered. Use the attached document as a foundation and include the following topics: Reality of change, Responsibility of action, economy and markets, pollution and biodiversity, sustainable agriculture, global perspective, action and responses."

   We appended a document with background information to provide context. The document contained Chapter: 3 Context from Mathys paper (Mathys, 2024) and all available slides form the expert inputs of the citizens' assembly. The output was then translated to German, French and Italian.

   Anchor statements for subcategories

   Original prompt:

   > "You are an experienced journalist. Write a summarizing paragraph about the future of Switzerland's food system concerning the topic of '[Insert Topic here]' on the basis of the attached file."

   We appended the same document with background information as above to provide context. The output was then translated to German, French and Italian.

   List of subcategories inserted:

| 1. | Reality of Change |
|----|-------------------|
| 2. | Responsibility of Action |
| 3. | Economy and Markets |
| 4. | Pollution and Biodiversity |
| 5. | Sustainable Agriculture |
| 6. | Global Perspective |
| 7. | Action and Responses |

# GitHub Link:

https://github.com/mooorice/automated-DRI-statement-generation

Since the data processed during the procedure amounts to almost 20GB and GitHub doesn't allow for large files to be uploaded, the project is uploaded without the corresponding data.

# AI Use Declaration

I hereby declare that I have used generative AI technologies in the preparation of this thesis solely within the permitted scope. The AI tools were used as writing assistants and coding support tools. At all times, I remained responsible for the content, critically reviewed all AI-generated suggestions, and ensured the correctness, coherence, and academic integrity of the work.

**AI Tools Used:**

**1. ChatGPT-4o and o1 (OpenAI)**

- **Purpose of use:**
    - Proofreading and improving grammar, style, formulation, readability, and fluency of the thesis text.
    - Rewording sentences to improve clarity while retaining the original meaning.
- **Example prompts used:**
    - *"Please improve the readability and academic style of the following paragraph without altering the meaning: [text]"*
    - *"Check this text for grammar and fluency errors and suggest stylistic improvements: [text]"*
    - *"Reformulate the following paragraph in clear, academic language: [text]"*
- **Verification measures:**
    - All suggestions were manually reviewed, checked against original content, and adjusted as needed.
    - I ensured that no new content was generated. Only existing text was refined.
    - I verified that the terminology and citations remained correct and appropriate.

**2. GPT-4o (OpenAI) and GPT-4o (o1 version)**

- **Purpose of use:**
    - Assistance with debugging code and optimizing code snippets used in development.
- **Example prompts used:**
    - *"How to solve this error: [error]"*
    - *"Write some code to [do x]."*
    - *"I'm having an issue with [x] what could be the problem?"*
- **Verification measures:**
    - All corrected code suggestions were tested and validated in my own coding environment.
    - I personally checked all outputs for flaws, misunderstandings and other issues.

**3. Claude 3.5 Sonnet (Anthropic)**

- **Purpose of use:**
    - Stylistic refinement and second-opinion checks on language quality and formulation.
    - Cross-checking paragraph coherence and balance in formulation.
- **Example prompts used:**
    - *"Help me improve the professional tone and flow of this academic section: [text]"*
    - *"Reformulate in an academic tone making sure to prevent redundancies and ensure stringency: [text]"*
- **Verification measures:**
    - AI suggestions were only used as stylistic input and manually verified.
    - I ensured no factual content or conclusions were altered or generated by the tool.

**Risk Minimization and Verification:**

- I manually reviewed and verified all AI-suggested text and code.
- I cross-checked all coding outcomes with manual testing.
- I validated all claims, arguments, citations and references without AI assistance.
- No part of the thesis content was generated entirely by AI. All content originated from my work and intellectual input.